\documentclass[aps, physrev, reprint, superscriptaddress]{revtex4-2}

\usepackage{silence}
\usepackage[utf8]{inputenc}
 \usepackage{xcolor}
\usepackage{amsmath}
\usepackage{amsfonts}
\usepackage{graphicx}
\usepackage{subcaption}
\usepackage{caption}
\usepackage{float}
\usepackage{appendix}
\usepackage{graphicx}% Include figure files
\usepackage{dcolumn}% Align table columns on decimal point
\usepackage{bm}% bold math
\usepackage{hyperref}
\usepackage{soul}

\begin{document}

%\preprint{APS/123-QED}

\title{Concatenating Binomial Codes with the Planar Code}% Force line breaks with \\

\author{Juliette Soule}
 \email{juliette.soule@sydney.edu.au}
 \affiliation{Centre for Engineered Quantum Systems, School of Physics, University of Sydney, Sydney, NSW 2006, Australia.}
\author{Andrew~C.~Doherty}%
 \email{andrew.doherty@sydney.edu.au}
 \affiliation{Centre for Engineered Quantum Systems, School of Physics, University of Sydney, Sydney, NSW 2006, Australia.}
 \author{Arne~L.~Grimsmo}%
 \affiliation{Centre for Engineered Quantum Systems, School of Physics, University of Sydney, Sydney, NSW 2006, Australia.}
\affiliation{AWS Center for Quantum Computing, Pasadena, CA 91125, USA}
        \affiliation{California Institute of Technology, Pasadena, CA 91125, USA}

\date{\today}% It is always \today, today,

\begin{abstract}
Rotation symmetric bosonic codes are an attractive encoding for qubits into oscillator degrees of freedom, particularly in superconducting qubit experiments. While these codes can tolerate considerable loss and dephasing, they will need to be combined with higher level codes to achieve large-scale devices. 
We investigate concatenating these codes with the planar code in a measurement-based scheme for fault-tolerant quantum computation.
We focus on binomial codes as the base level encoding, and estimate break-even points for such encodings under loss for various types of measurement protocol. These codes are more resistant to photon loss errors, but require both higher mean photon numbers and higher phase resolution for gate operations and measurements. We find that it is necessary to implement adaptive phase measurements, maximum likelihood quantum state inference, and weighted minimum weight decoding to obtain good performance for a planar code using binomial code qubits.

\end{abstract}
\maketitle

\section{Introduction}
One of the major challenges in building a quantum computer is implementing error correction. Error correction, the encoding of quantum information to protect it against environmental noise, is crucial in order to perform large-scale calculations. In this work we are interested in the performance of bosonic codes in which qubits are encoded into the infinite-dimensional Hilbert space of a harmonic oscillator.  Bosonic codes can provide protection against typical sources of noise such as loss and dephasing~\cite{bosonic1,bosonic2,cochrane1999macroscopically} and can be implemented in a variety of physical architectures, including electromagnetic modes, for example microwave modes controlled by superconducting qubits~\cite{catexperiment, binomexperiment, CampagneIbarcq2020}, or mechanical degrees of freedom such as trapped ion motional modes \cite{fluhmann2019encoding, de2022error}. 

Experiments in the context of superconducting qubits have shown that bosonic codes are amongst the most promising approaches to practical quantum computing in that system. Specifically it has been possible to operate these experiments beyond the memory break even point at which the lifetime of an encoded qubit equals that of an unencoded qubit~\cite{catexperiment, gkpbreakeven}.
In addition it is possible to perform logical gates and fault tolerant measurements in such experiments~\cite{ binomexperiment, ftcat, ftbosonic, ftbosonic2, ftbosonic3,SCbosonic, 100PhotonCat}.

No matter how good bosonic codes become, however, there will be residual errors. Large scale quantum computation will require bosonic codes to be concatenated with a higher level code, such as a surface code~\cite{2002, surfacecode}, that allows fault tolerant quantum computation~\cite{bosonic2}. The bosonic code must be able to suppress the noise below the fault tolerance threshold of the higher level code. For the so-called \emph{GKP codes}~\cite{bosonic2} there have been studies of how to achieve this~\cite{surfaceGKP1,surfaceGKP2,surfaceGKP3} but for widely used codes such as the so-called \emph{cat codes}~\cite{catkerr, PhysRevLett.111.120501, Mirrahimi_2014} and \emph{binomial codes}~\cite{Michael_2016} there is not a quantitative understanding of the performance required for large-scale computation. 

%enabling the overall scheme to be more hardware efficient \cite{PhysRevLett.100.030503, Mirrahimi_2014, PhysRevX.6.031006, Puri_2019, Li_2021}. 

%Bosonic codes were first introduced by Gottesman, Kitaev and Preskill \cite{bosonic2}. The eponymous GKP codes possess discrete translation symmetry in phase space which enables detection of small shifts in phase space.

In this work we describe and implement simulations to determine the performance of an architecture that concatenates the surface code with a class of bosonic codes that Grimsmo and collaborators  termed \emph{rotational symmetric codes}~\cite{arne}. They provided a unified analysis of these codes, which include both cat codes and binomial codes, as well as a theoretical scheme for logical gates, measurements, and state preparation~\cite{arne}. Our approach uses this gate set to study the performance of a specific choice of higher level code and architecture for quantum computation. 

%are similar to GKP codes but instead characterised by their discrete rotational symmetry in phase space. In general, increasing the order of the rotational symmetry will increase the tolerance of the code to photon gain and loss errors, whilst reducing tolerance to dephasing errors.

The gate set proposed in~\cite{arne} suggests that 
a natural approach to realising fault tolerant compuation with concatenated bosonic codes is Measurement Based Quantum Computation (MBQC) with code foliation \cite{foliationog, foliatedcss,,benpaper}. In particular, for rotation symmetric codes the natural ``easy'' operations include a $\bar{C}_{Z} = \text{diag}(1,1,1,-1)$ gate, realized with a cross-Kerr interaction between two modes, and X-basis measurement, realized as a phase measurement~\cite{arne}. Together with state preparation of $|+\rangle$ states, this forms the the basis for MBQC.
%Consider the hardware platform of superconducting qubits, a leading method for realising bosonic codes. In this platform, it is experimentally feasible to engineer a cross-Kerr interaction between encoded qubits, which realises a controlled-rotation entagling gate. This can act as a logical $\bar{C}_{Z} = diag(1,1,1,-1)$ gate \cite{arne}. MBQC schemes are heavily dependent upon $\bar{C}_{Z}$ gates along with X basis measurements, both of which are experimentally feasible for bosonic encoded qubits.
These considerations motivate our scheme which uses bosonic encoded qubits, in particular binomial qubits, as the 'physical qubits' of a 2D surface code, realised by MBQC.

The aim of this work is to conduct a preliminary investigation of this quantum computation scheme, taking into account a realistic model of photon loss which is the primary source of noise for bosonic codes. Furthermore, the X basis measurements that we require for our scheme will be realised as phase measurements of a bosonic mode. The phase measurement itself, as well as the procedure for inferring the qubit state, are imperfect processes which will introduce inaccuracy to our scheme. We will compare different methods of phase measurement such as heterodyne and adaptive homodyne measurement \cite{ahdmain}, combined with different procedures for qubit state inference. %We aim to obtain a realistic picture of the impact of photon loss, and imperfect measurement on bosonic codes in a concatenated MBQC architecture.

%Using this scheme, w
We obtain threshold values for varying orders of discrete rotational symmetry. We find that whilst increasing the order of discrete rotational symmetry generally improves the threshold value, thresholds for binomial encoded qubits tend to be lower than for the trivial Fock space encoding,  $|0\rangle_{L} = |0\rangle, |1\rangle_{L} = |1\rangle$, when using the most naive measurement and qubit state inference techniques. By using more sophisticated measurements and state inference methods as well as incorporating the soft information from measurements into decoding we are able to find schemes that beat the trivial encoding when qubit state measurement error for the trivial encoding is not too small. 

The point at which the lifetime of a bosonic encoded qubit outperforms the trivial encoding is referred to as break even \cite{arnerecent}. The competition between the bosonic codes and the trivial encoding arises because the rotationally symmetric codes achieve tolerance to loss errors at the expense of increased Fock number. In comparison, the trivial encoding has no intrinsic robustness to loss but has a very low average number. We find that optimal measurement and state inference is required for the rotational symmetric codes to outperform the trivial encoding. Finally, we examine the sub-threshold performance of binomial codes and find that below $50\%$ of threshold the loss tolerance of a binomial code is more beneficial and codes with increased rotational symmetry  perform better than the trivial code and better than codes with reduced symmetry. 

This paper is structured as follows. In Sec. \ref{setup} we review rotation symmetric bosonic codes and measurement based quantum computation. In Sec. \ref{1danalysis} we introduce the measurement models and qubit state inference techniques used to realise the computational scheme. Sec. \ref{1dresults} then quantifies the performance of these methods. In Sec. \ref{2danalysis} we introduce methods used for the 2D planar code, including an X basis Pauli twirling approximation and modified MWPM decoding algorithm. Sec. \ref{mainresults} compares the code thresholds obtained using these techniques, as well as examining the subthreshold scaling of the code. Finally, in Sec. \ref{end} we review our findings and highlight the main conclusions.

\section{Setup and Description of Cluster State Model}\label{setup}
In this section we will define the various elements of the error correction scheme we investigate. This scheme involves encoding qubits in bosonic modes using a rotational symmetric encoding, specifically the binomial encoding. These encoded qubits are then entangled using ${C}_{Z}$ gates to form a specific two dimensional many-body entangled resource state, also known as a cluster state, which maps onto the two dimensional surface code \cite{foliationog}. In this work, we will restrict to the 2D surface code. The rotation symmetric bosonic (RSB) encoded qubits are then measured and subject to error correction in order to enact the identity gate on the code. Note that the scheme could be extended to non-trivial gates, however this will not be considered in this work.
Though the 2D surface code is not fault tolerant, it is a more straightforward setting in which we conduct a preliminary investigation into our concatenated scheme.

\subsection{Rotation Symmetric Bosonic Codes} \label{RSB}
Encoding a qubit in a single bosonic mode involves defining a two dimensional subspace of the mode's Hilbert space as the span of the logical codewords. The remaining space can be used to detect and correct errors. The most prevalent error sources to which bosonic modes are subject are loss and dephasing. Rotation-symmetric  encodings have been tailored specifically to be able to correct loss and can exactly correct up to a certain order of photon loss. This property is tightly connected to the discrete rotational symmetry of these codes.

\begin{figure}
    \centering
    \includegraphics[trim=0cm 13cm 7cm 5cm,clip,scale=0.5]{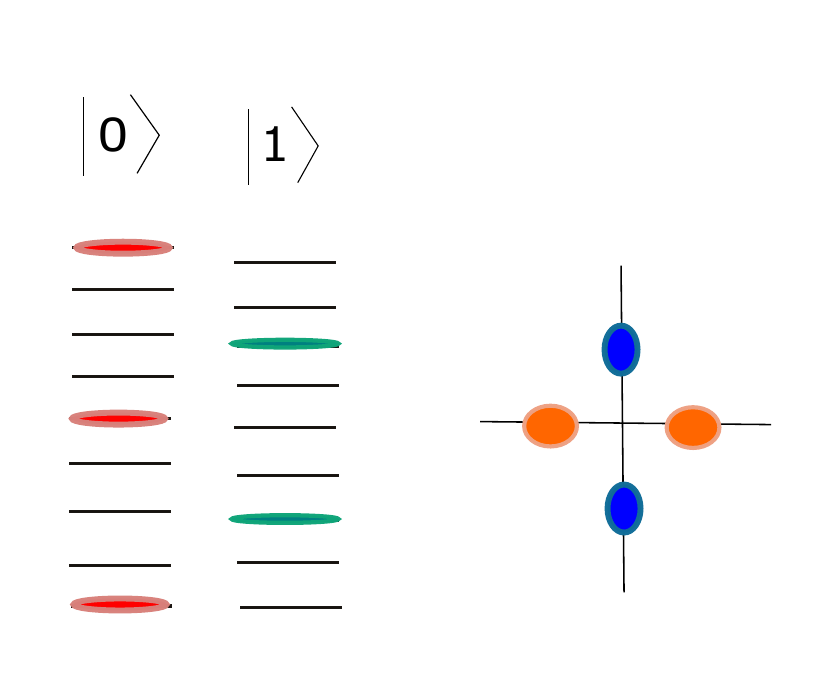}
    \caption{Fock and phase space structure of an $N = 2$ RSB code. Fock states supporting $|0\rangle$ are offset by $N$ from those supporting $|1\rangle_{N}$, whilst Fock states supporting either $|0\rangle_{N}$ or $|1\rangle_{N}$ are offset by $2N.$ $|\pm\rangle_{N}$ states (orange and blue respectively) are separated by $d_{\theta} = \pi/N$ in phase space.} 
    \label{fig:rsbfockspace}
\end{figure}

Rotation-Symmetric Bosonic (RSB) codes~\cite{arne} have logical $Z$-operators of the form
\begin{align}
    \hat{Z}_{N} &= e^{i\pi\hat{n}/N},
\end{align}
where $\hat{n} = \hat{a}^{\dagger}\hat{a}$ is the Fock space number operator, and $\hat{a},\hat{a}^{\dagger}$ are the Fock space annihilation and creation operators, respectively. Such codes have discrete rotational symmetry of order $N$ as the projector onto the codespace
commutes with the discrete rotation operator
\begin{align}
    \hat{R}_{N} &= e^{i2\pi\hat{n}/N}.
\end{align}
Indeed, $\hat{R}_{2N} = \hat{Z}_{N}.$ The choice of the logical Z operator determines the form of the logical codewords as finite rotated superpositions of a ``primitive'' state \cite{arne} $|\theta\rangle$, where different types of rotation symmetric codes, such as cat and binomial codes, differ only in the choice of primitive. Explicitly, for $\mu\in\{0,1\}$
\begin{align}
    |\mu\rangle_{N} &= \frac{1}{\sqrt{\mathcal{N}_{\mu}}}\sum_{m=0}^{2N-1}(-1)^{\mu m}e^{i\pi m \hat{n}/N}|\theta\rangle.
\end{align}
The $\hat{Z}_{N}$ operator also enforces a specific Fock grid structure of the logical codewords. In particular, $|\mu\rangle_{N}$ for $\mu\in\{0,1\}$ has support on every $|(2k+\mu)N\rangle$ Fock state, as per Fig. \ref{fig:rsbfockspace}. This structure results in a Fock grid distance $d_{n} = N$ between logical zero and logical one states.

Dual basis codewords are defined in the usual way 
\begin{align}
    |\pm\rangle_{N} &= \frac{1}{\sqrt{2}}(|0\rangle_{N}\pm|1\rangle_{N}).
\end{align}
In contrast to primal codewords, they have support on every Fock state $|kN\rangle.$ They are separated by a distance in phase space $d_{\theta} = \pi/N.$ Note that $d_{n}\propto N$ whilst $d_{\theta} \propto \frac{1}{N}.$ Therefore, whilst increasing $N$ is advantageous for the resilience of $|0\rangle_{N},|1\rangle_{N}$ codewords to loss and gain errors, it is detrimental to the capacity of $|\pm\rangle_{N}$ codewords to tolerate dephasing errors.

Binomial codes are so called due to the binomial coefficient weighting of their Fock state coefficients. In particular
\begin{align}
    |\mu_{N}\rangle_{\mathrm{bin}}&=\sum_{k=0}^{\left\lceil{\frac{K}{2}}\right\rceil-\mu}\sqrt{\frac{1}{2^{K-1}}\binom{K}{2k+\mu}}|(2k+\mu)N\rangle
\end{align}
where the parameter $K$ relates to the number of loss and dephasing errors that are correctable by the code, as detailed in \cite{Michael_2016}.

 In the following we will omit the $N$ subscript in our notation and simply use $|\psi\rangle$ to denote an RSB qubit encoded in the state $\psi.$ When referring to `logical' operations and states, we will mean operations and states at the level of the surface code.
 
 We study a quantum computing scheme which requires the preparation of RSB $|+\rangle$ states, destructive X basis measurement, and an RSB $C_{Z}$ gate. These three elements are sufficient to enact all Clifford gates \cite{arne}. Universality may be achieved for this scheme by injecting magic states but non-trivial gates are beyond the scope of this work.

\subsubsection{RSB $C_{Z}$ Gate}
As discussed above the only logical gate we need to enact on RSB encoded qubits in our computational scheme is the controlled phase $C_{Z}$ gate.
The $C_{Z}$ gate may be realised by a controlled rotation at the physical level generated by a cross-Kerr interaction between the modes \cite{catkerr, catkerr2}
\begin{align}
    H_{\textsc{CROT}_{ij}} &= -\Omega \hat{n}_{i}\otimes\hat{n}_{j} \nonumber \\
    \textsc{CROT}_{ij} &= e^{-iH_{\textsc{CROT}_{ij}}t_{\rm gate}} \nonumber \\
    &= e^{\frac{i\pi\hat{n}_{i}\otimes\hat{n}_{j}}{NM}}
\end{align}
where modes $i,j$ involve RSB qubits with discrete rotational symmetry of order $N,M$ respectively. Loss is the dominant imperfection in our model and if the cross-Kerr gate is slow the effect of loss is exacerbated. Both the strength of the loss and the time required for the gate are captured by the parameter $\gamma=\kappa t_{gate}$ which we will use to characterize the noise level. The gate time is chosen such that $\Omega t_{\rm gate}=\pi/NM$ thus $\gamma$ is proportional to the photon loss rate and inversely proportional to the cross-Kerr nonlinearity. In all that follows we only consider $\textsc{CROT}$ gates between identical encodings, so $N = M.$ 

Note that low cross-Kerr nonlinearity in practical devices might lead to a different coupling non-linearity being preferred, such as a SNAIL~\cite{Frattini_2017}. In this case as well the effect of loss will depend on the per photon loss probability during a gate. So while the details of code performance would differ, we expect qualitatively similar results for such a gate.

The other RSB operation we need is a destructive single qubit X measurement, which we investigate in Sec.\ref{singlequbitmsmt}.

\subsection{Noise Model}\label{noise}
A primary noise source plaguing bosonic modes is photon loss. We consider a model in which photon loss occurs during the implementation of each $\textsc{CROT}$ gate. 

\subsubsection{Photon Loss Noise}
To describe the noise due to photon loss, we use the framework of quantum trajectory theory~\cite{wiseman2009quantum} to analyse a master equation describing losses occurring on each qubit during the implementation of the $\textsc{CROT}$ gate. The master equation describing the combined effects of the gate and loss is
\begin{align}
    \dot{\rho} =-i[H_{\textsc{CROT}_{12}},\rho] +\kappa\sum_{i=1}^{2}\mathcal{D}[\hat{a_i}]\rho
\end{align}
where $\hat{a}_{1}=\hat{a}\otimes\mathbb{I},\hat{a}_{2}=\mathbb{I}\otimes \hat{a}$, $\mathcal{D}[L]\rho = L\rho L^{\dagger}-L^{\dagger}L\rho/2-\rho L^{\dagger}L/2$ is the Lindblad superoperator. We will describe the effects of loss using the parameter $\gamma = \kappa t_{\rm gate}$, 

In the trajectory approach the density matrix can be written as a sum over the number of photon emission events that occur during the gate. Each term involves an integral over the emission time of the photons. As described in more detail in Appendix \ref{quanttraj} we commute the loss past the $\textsc{CROT}$ gate in each such term so that we are left with an effective error operator acting on each qubit. 

We will introduce a notation that can be generalized to the case of $M$ modes. Suppose that the times for which losses occur on mode 1 are listed in a vector $\mathbf{t}_1$. We will suppose that these emission times are arranged in time order such that for times $t_m,t_{m+1}\in \mathbf{t}_1$ we have $t_{m+1}>t_m$. Similarly emission times for mode 2 are $\mathbf{t}_2$. We will sometimes need to refer to the emission times for all modes, and we do this by defining an array of emission time vectors
$\mathbf{t}=\{\mathbf{t}_1,\mathbf{t}_2\}.$ The number of photon emissions for mode $a$ is $j_{a}=|\mathbf{t}_{a}|$ the number of entries in the vector $\mathbf{t}_{a}$. The total number of photon emissions $k$ is the sum of the $j_a$, so for two modes we have $j_1+j_2=k$.  Then the noisy $\textsc{CROT}$ gate with a fixed number of photon emissions can be expressed as
\begin{align}
\widetilde{\textsc{CROT}}_{12,\mathbf{t}}
    &= \hat{E}_{1,{\mathbf{t}}}\hat{E}_{2,{\mathbf{t}}}\textsc{CROT}_{12}
\end{align}
where $\hat{E}_{{\mathbf{t}}},\hat{E}_{{\mathbf{t}}}$ are error operators acting on qubits 1,2 respectively,
%that correspond to $j_1$ and $j_2$ photon loss events at times $\mathbf t_1,\mathbf t_2$ and are
given by
\begin{subequations}\label{croterror}
\begin{align}
    \hat{E}_{1,{\mathbf{t}}} =& \sqrt{\kappa}^{j_1}e^{\kappa \tau(\mathbf{t}_1)/2}e^{i\Omega \tau(\mathbf{t}_2)\hat{n}_1}\hat{a}_1^{j_1} e^{-\kappa t_{\rm gate}\hat{n}_{1}/2}
   \\
   \hat{E}_{2,{\mathbf{t}}}=& \sqrt{\kappa}^{j_2}e^{\kappa \tau(\mathbf{t}_2)/2}e^{i\Omega \tau (\mathbf{t}_1)\hat{n}_2}\hat{a}_2^{j_2} e^{-\kappa t_{\rm gate}\hat{n}_{2}/2}.
\end{align}
\end{subequations} 
This expression is derived in the Appendix \ref{quanttraj}, specifically Eq. \ref{identityequation}. The effective delay parameters $\tau(\mathbf{t}_a)$ appearing in these expressions are given by
\begin{equation}\tau(\mathbf{t}_a)= j_at_{\rm gate}-\sum_{t_m\in \mathbf{t}_a}t_m.\end{equation}

The various factors in $\hat{E}_{1,\mathbf{t}}$ have a natural interpretation. $\hat{E}_{1,{\mathbf{t}}} $ removes $j_1$ photons from mode $1$, hence the factor $\hat{a}_1^{j_1}$. The probability of an event with this number of photon emissions, and the resulting conditioning of the wavefunction, are described by the constant factor $\sqrt{\kappa}^{j_1}\exp(\kappa \tau(\mathbf{t}_1)/2)$ and the non-unitary operator $\exp(-\kappa t_{\rm gate}\hat{n}_{1}/2)$. Finally the action of the gate during the photon loss means that photon loss events on the neighbouring mode $2$ lead to a phase shift on mode 1. This correlated noise process is described by the unitary factor $\exp[i\Omega \tau(\mathbf{t}_2)\hat{n}_1]$. It has very significant consequences for the performance of error correction as we will see.

Turning now to the multimode case where we will need to perform many simultaneous $\textsc{CROT}$ gates. Specifically the gates to be performed are represented by a graph $G = (V,E)$ where the vertices $V$ represent qubits and the edges $E$ represent the location of the gates.  We will also use the notation of neighbourhoods - that is, a qubit $i$ is said to be in the neighbourhood of qubit $j$, $i\in\mathcal{N}(j)$, if qubits $i,j$ share an edge. We consider applying all $\textsc{CROT}$ gates simultaneously. Using the same notations as for the two-qubit case, we can express the net noise operator for a qubit $a$ as
\begin{align}
\hat{E}_{a,\mathbf{t}} = \sqrt{\kappa}^{j_{a}}e^{\kappa \tau(\mathbf{t}_a)/2}\left(\prod_{b\in\mathcal{N}(a)}e^{i\Omega \tau(\mathbf{t}_b)\hat{n}_a}\right)\hat{a}_{a}^{j_{a}}e^{-\kappa t_{\rm gate}\hat{n}_{a}/2}.
\end{align}
Note that the mode $a$ acquires a phase shift for a photon emission on any neighbouring qubit.

The overall effect of the noise for a given set of photon emission times $\mathbf{t}$ is given by the operator \begin{equation}\tilde{U}_{\mathbf{t}}=\left(\prod_{a=1}^M\hat{E}_{a,\mathbf{t}}\right)\prod_{e\in E}\textsc{CROT}_{e_1,e_2}\end{equation} This equation, derived in the Appendix as equation~\ref{identityequation}, expresses the noise as an ideal gate followed by a modified photon loss noise operator. We can think of the overall effect of this noise as defining a noise operator that acts on the ideal cluster state as follows
\begin{align}\label{eq10}
     \tilde{\Gamma}_{\mathbf{t}}(\rho_{CS})
&=\tilde{U}_{\mathbf{t}}\left(|+\rangle\langle+|\right)^{\otimes M}\tilde{U}^\dagger_{\mathbf{t}}
\end{align}
where 
\begin{align}\label{noise}
   \rho_{CS} = U\left(|+\rangle\langle+|\right)^{\otimes M}U^\dagger
   \end{align}
and $U$ is the ideal gate with $\kappa=0$ and no photon loss events.

 This leaves us to determine how to correctly sample a set of emission times $\mathbf{t}$.  We show in Appendix \ref{quanttraj} that the probability distribution for photon emission times is unnaffected by the application of the $\textsc{CROT}$ gates. This greatly simplifies our simulations since the photon emission times can be determined independently for each qubit according to a probability distribution that is independent of the particular choice of $G$. The details of the sampling of emissions times $\mathbf{t}$ are givein in Appendix~\ref{dist}.

%%Our model uses rotation symmetric bosonic encoded qubits as the 'physical qubits' of a surface code, where the entangling gate is the logical $C_{Z} = CROT$. It is realised by using measurement based quantum computation (MBQC) by foliation. Consider some number $n$ of unentangled, RSB encoded qubits, where we wish to prepare an $n-1\times n-1$ planar surface code. We prepare the cluster state by encoding the first row of qubits in some logical state $|\psi\rangle\langle\psi|_{L}$. All other qubits are prepared in the $|+\rangle\langle+|_{L}$ state. We then concantenate each of the qubits in the initial logical state into the codespace of a 1D cluster state by entangling each of these qubits with a 1D cluster state. Different 1D cluster states are then entangled using ancilla qubits to concatenate the codespace of the 1D cluster states into another code. In our case, adjacent 1D cluster states are entangled to realise the 2D planar code. All the qubits of the cluster are then measured in the X basis, which propagates the logical state $|\psi\rangle\langle\psi|_{L}$ from the first layer to the final layer of the cluster. This realises an identity gate on the logical state. For a detailed explanation of this construction see \cite{benpaper}.
%%

\subsection{Measurement Based Quantum Computation}\label{mbqc}
Measurement Based Quantum Computation (MBQC) \cite{2007, foliationog}  is an alternative to the circuit-based model of quantum computing. In MBQC, computation is performed by preparing an entangled many-body resource state upon which single qubit measurements are performed, realising quantum gates. The resource state, often referred to as a cluster state, is entangled exclusively using ${C}_{Z}$ gates. The single qubit measurements on the cluster state effectively consume the qubits and drive the computation. 

In this work we will use MBQC to implement the 2D surface code \cite{Raussendorf_2007}. Whilst the 2D surface code is not fault-tolerant, it provides a simpler setting to explore the implications of realistic noise and measurement on code performance than a full 3D simulation. 

In this section we will first describe some preliminary notation used for cluster states, before detailing the construction of 1D cluster states. We will then explain how 1D cluster states can be entangled to realise the 2D surface code as a foliation of the repetition code.

\subsubsection{Cluster States}
A cluster state can be associated to a graph $G = (V,E)$ where vertices represent qubits. Two vertices share an edge if the two qubits of the cluster state are entangled via a ${C}_{Z}$ gate.
%In this case we associate qubits $e_1,e_2$ to an edge $e.$
In our model, each vertex of the graph represents an RSB qubit, and vertices share an edge if the qubit pair is entangled via a $\textsc{CROT}$ gate. The cluster state provides the resource for the quantum computation. Computation will be carried out by single qubit measurements, which will be elaborated upon in Sec. \ref{mbqc}.

\subsubsection{1D Cluster State Description}\label{1Dcluster}
We begin by constructing a 1D cluster state and using it to teleport a qubit along the line. For a length $M$ 1D cluster state, we prepare a single qubit in an arbitrary state $|\psi\rangle,$ and the remaining qubits in the $|+\rangle$ state. We then entangle qubits in a line by applying $C_{Z}$ between adjacent qubits. We then measure each qubit in the X basis, which teleports the logical state from the first qubit, down the chain to the final qubit.

We can understand the 1D cluster state in terms of its stabilisers, as in \cite{biasedCS} for example. Initially, the logical operators describing the 1D chain will simply be $X_{L} = X_{1}, Z_{L} = Z_{1}$ and before the $C_Z$ gates the stabiliser group is given by $S = \{X_{1},X_{2},...,X_{n}\}.$ After applying ${C}_{Z}$ gates, operators transform as
\begin{align}
    Z_{i}&\rightarrow Z_{i}\nonumber \\
    X_{i}&\rightarrow X_{i}\prod_{j\in\mathcal{N}_{i}}Z_{j}
\end{align}
where $\mathcal{N}_{i}$ denotes all qubits which are entangled to qubit $i$ by a ${C}_{Z}$ gate. This causes logicals and stabilisers to transform as
\begin{align}
    Z_{L} &= Z_{1}\nonumber\\
    X_{L} &= X_{1}Z_{2}\nonumber\\
    S &= \{X_{1}Z_{2},Z_{1}X_{2}Z_{3},...,Z_{M-1}X_{M}\}.
\end{align}
At the conclusion of the $X$ measurements the first $M-1$ qubits are left in $X$ eigenstates and by multiplication with stabilizers we find that the logicals become $Z_{L} = Z_{M}$ and $X_L=X_M$ reflecting teleportation along the chain.

It is useful to consider how loss at any given location on the chain affects the performance of the scheme.  Loss on qubit $i$ propagates through the controlled rotation (${C}_{Z}$) to act as dephasing on qubits $i\pm1$. In our simulations of this teleportation scheme we assess performance by assuming that the final qubit in the chain has no noise and  applying a noise-free recovery to assign a fidelity to the teleportation procedure. 

\subsubsection{2D Surface Code}
\begin{figure}[h]
    \centering
    \includegraphics[trim=4cm 13cm 7cm 4cm,clip,scale=0.3]{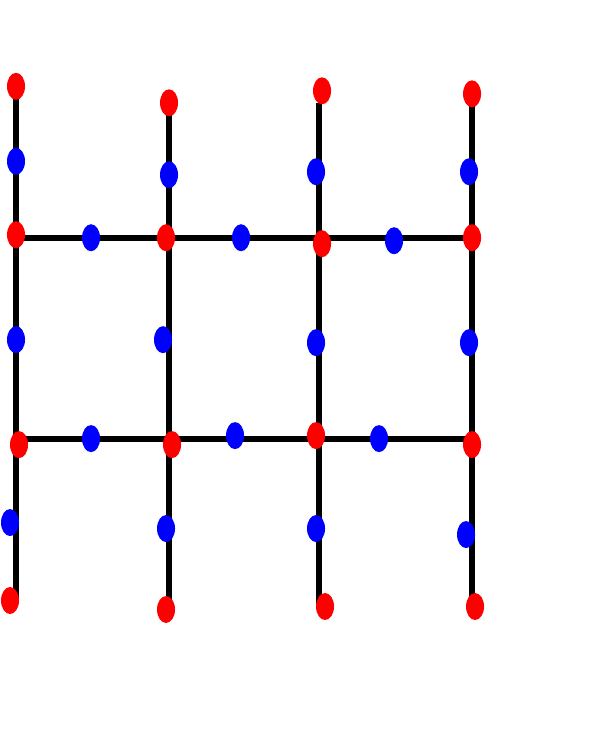}
    \caption{1D cluster states foliated into the 2D planar code. Blue qubits are primal, red qubits are dual.}
    \label{fig:2Dplanar}
\end{figure}

\begin{figure}[h]
\begin{center}
\includegraphics[scale=1.3]{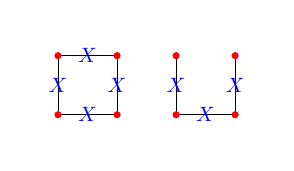}
\caption{Primal plaquette stabilisers for the 2D planar code for the bulk and rough boundary, respectively. Blue 'X' indicates X measurements of primal qubits that contribute to the stabiliser. Red circles are dual qubits which do not contribute to the stabiliser measurement.}\label{fig:Zstabilisers}
\end{center}
\end{figure}

\begin{figure}
\begin{center}
\includegraphics[trim=2cm 13cm 4cm 4cm,clip,scale=0.3]{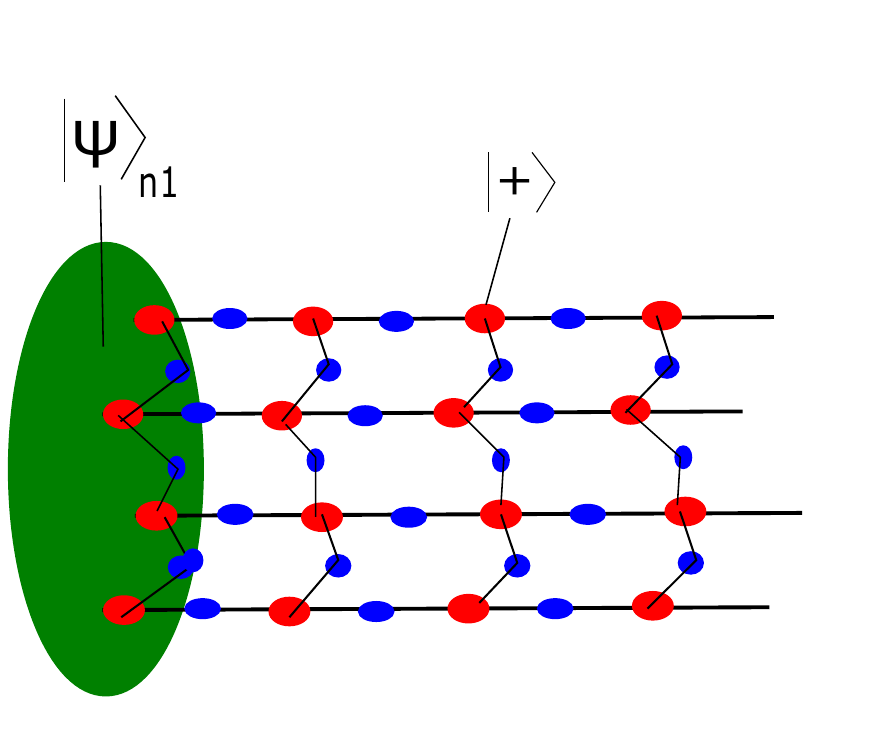}
\caption{A cluster state used to implement the 2D surface code using MBQC. The green oval represents $|\psi\rangle_{L},$ the initial logical state on the first layer of qubits. The remaining qubits are initialised in the $|+\rangle_{L}$ state.}\label{2dcluster}
\end{center}
\end{figure}

In order to construct the 2D surface code in Fig. \ref{fig:2Dplanar}, we start with parallel 1D cluster states, prepared as before and entangle every odd (primal) qubit of each 1D cluster with the corresponding primal qubit of its neighbouring 1D cluster state, via a dual qubit prepared in the $|+\rangle$ state. Every second qubit of the 1D cluster state is also a dual qubit. The resulting cluster state is depicted in Fig. \ref{2dcluster}. 
Measuring every qubit in the X basis realises a measurement of parity checks of primal stabilisers given by a product of $X$ on qubits around a primal plaquette. A primal plaquette stabiliser checks for $Z$ type errors on that plaquette.
More explicitly, consider the stabilisers of the cluster state.
For a plaquette $p$ of the cluster state, the associated stabiliser is $S_{p} = \prod_{e\in p}X_{e}$, where $e$ denote the primal qubits represented by the edges of the given plaquette as in Fig. \ref{fig:Zstabilisers}. To teleport logical information across the 2D foliated state, we measure both primal and dual qubits in the X basis. This also provides stabiliser outcomes, from which we extract the error syndrome. 

Having obtained the error syndrome, error correction can be carried out in the usual way \cite{2002, surfacecode}.  We use a minimum weight perfect matching (MWPM) algorithm to match up pairs of violated stabilisers in the error syndrome in a way that minimises the total distance between pairs of violated stabilisers. The paths between violated stabilisers given by MWPM determine which qubits have experienced errors and therefore give the Pauli correction to which the final logical state with be subject. After the correction has been applied, the simulation can perform a hypothetical logical measurement that determines if the correction has succeeded, or if it has failed and the code has suffered a logical error.

The model we simulate can either be viewed as the foliated 1D repetition code, or equivalently a single 2D timeslice of plaquette stabilizers of the surface code. The noise model we adopt for this scheme includes photon loss occurring during gates and measurement errors on single qubit X measurements, the latter of which arise naturally from the realistic measurement model employed. In the commonly used phenomenological error models for the surface code, both these noise sources map to gate noise rather than phenomenological measurement noise, so that we are still able to observe an error correction threshold for this scheme. Note however that there is only a single logical operator that can be corrected. Therefore while it is possible to learn a lot about the interplay between loss and measurement errors in this code concatenation scheme, a larger scale simulation corresponding to a $2+1D$ surface code simulation would still be desirable.

%%%
%Consider a foliated cluster state of $p$ total qubits. Let the bottom layer of the cluster be comprised of $n_{1}$ qubits, and let $n_{2} = p - n_{1}$. Suppose that the bottom layer of the cluster is initialised in the logical state $|\psi\rangle\langle\psi|_{n_{1}}$, and that the remaining cluster qubits are initialised in the $|+\rangle\langle+|_{L}$ state. Suppose that cluster qubits are entangled according to the prescription above. Denote qubits $i,j$ as entangled if $j\in\mathcal{N}(i)$. Letting $\tilde{\Gamma}$ denote the noise channel having been commuted past the $CROT$ gates, we may express the full cluster state as
%\begin{align}
%    \rho_{CS} &= \prod_{i=1,j\in\mathcal{N}(i)}^{p}\tilde{\Gamma}(CROT_{ij}(|\psi\rangle\langle\psi|_{n_{1}}\otimes|+\rangle\langle+|^{\otimes n_{2}})CROT^{\dagger}_{ij}).
%\end{align}
%%%%

\section{Analysis and Numerics}\label{1danalysis}

%\subsection{1D Cluster State Methods} 
\subsection{Single Qubit Phase Measurement} \label{singlequbitmsmt}
We require X basis measurements of individual qubits for our quantum computation scheme. This measurement translates to a phase-estimation problem based on measurements of a single bosonic mode. Feasible methods for conducting phase measurements are heterodyne measurement and adaptive homodyne (AHD) measurement~\cite{ahdmain}. We compare the performance of heterodyne and AHD measurement to an ideal canonical phase measurement which would be the optimal choice of measurement if it could be implemented.

All phase measurements can be defined using the general positive operator-valued measurement (POVM)~\cite{ahdmain} 
\begin{align}
    \hat{F}(\phi) &= \frac{1}{2\pi}\sum_{n,m = 0}^{\infty}e^{i\phi(m-n)}H_{mn}|m\rangle\langle n| 
\label{eqn:phasePOVM}
\end{align}
where $\phi\in[0,2\pi)$ is the measurement outcome and H is a Hermitian matrix with real positive entries and $H_{mm}=1$ for all $m$ that is defined explicitly in Appendix \ref{ahdappendix}. The quality of the phase measurement depends on the choice of $H$.

\subsubsection{Canonical Phase Measurement}
The ideal realisation of a phase measurement of a bosonic mode is simply the projector onto an (unnormalised) phase eigenstate \cite{idealmsmt}. 
This measurement model is very hard to implement, however we include it in simulations as a benchmark of the best possible measurement to which we can compare the other schemes implemented. It has POVM elements
\begin{align}
    \hat{F}_{\mathrm{c}}(\phi)&=\frac{1}{2\pi}|\phi\rangle\langle\phi|
\end{align}
where the phase eigenstate is $|\phi\rangle = \sum_{n=0}^{\infty}e^{i \hat{n}\phi}|n\rangle.$
\subsubsection{Heterodyne Measurement}
Heterodyne measurement involves simultaneous homodyne measurements of orthogonal quadratures of a bosonic mode, resulting in adding noise to both quadratures. Heterodyne measurement projects the qubit onto the coherent state $|\alpha\rangle\langle\alpha|$. We can obtain then obtain the phase as $\mathrm{arg}(\alpha) = \phi.$ The POVM for heterodyne measurement with outcome $\alpha$ is
\begin{align}
    \hat{F}_{\mathrm{coh}}(\alpha)&=\frac{1}{\pi}|\alpha\rangle\langle\alpha|.
\end{align}
and \begin{equation}\hat{F}_{\mathrm{h}}(\phi)=\int_0^\infty \hat{F}_{\mathrm{coh}}(re^{i\phi})dr.\end{equation}

\subsubsection{Adaptive Homodyne Measurement}
Adaptive Homodyne Measurement (AHD) \cite{ahdmain} is a better performing alternative to heterodyne measurement. Ordinary homodyne measurement involves the measurement of a single quadrature of the harmonic oscillator mode. It has lower noise than heterodyne measurement but cannot determine the phase of the bosonic mode. In AHD measurement, however, the phase $\theta$ of the local oscillator field is continuously updated based on prior measurements. This rotates the quadrature that is being measured and the scheme is designed to lock in of the phase of the field. Various adaptive schemes are possible we use the Mark II scheme from~\cite{ahdmain}.  The AHD POVM elements are determined by a specific choice of the matrix $H_{mn}$ that appears in Eq.~\ref{eqn:phasePOVM}, which is defined in Appendix \ref{ahdappendix}.

A detailed discussion of both AHD and canonical phase measurement schemes can be found in~\cite{ahdmain}. An experimental realisation of AHD measurement is described in~\cite{ahdexpt}.

\subsubsection{Measurement Error}\label{msmterrorsection}
In our scheme phase measurements are used to implement $X$ basis measurements of a RSB code. The phase measurement must be able to resolve angles less than $\pi/N$ in order to distinguish qubit states. Thus for a fixed phase measurement it becomes harder to perform qubit measurement as $N$ increases. Neither AHD nor heterodyne measurements provide ideal projective qubit measurements. Both have inherent measurement error which is dependent upon the mean photon number, in addition to the requirement to resolve angles of order $\pi/N$.  

Since in this setting we just require the ability to distinguish the two qubit $X$-eigenstates based on phase measurement, we will use a Qubit State Inference (QSI) procedure to achieve this.   If the inferred value of the qubit state differs from `actual' state of the qubit determined by our simulation then that constitutes a qubit measurement error. As we will explain in greater detail below this QSI procedure can be chosen to partial compensate from the phase errors that we have seen occur correlated to photon losses on neighbouring qubits.

\begin{figure}
\includegraphics[scale=0.4]{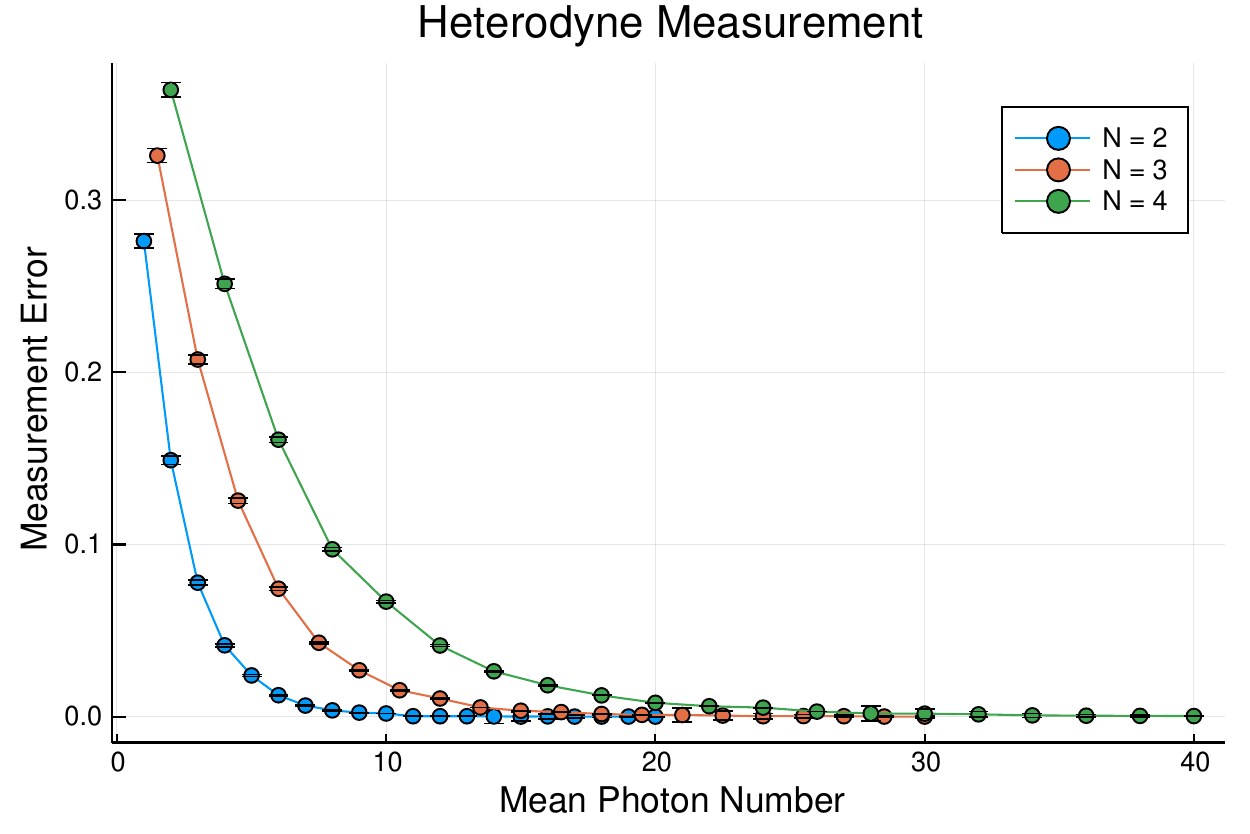}
\caption{Measurement Error vs $\bar{n}=1/2\cdot KN$ for binomial code with different $N$ values using heterodyne measurement.}
\label{fig:hetmsmterror}
\end{figure}

\begin{figure}
\includegraphics[scale=0.4]{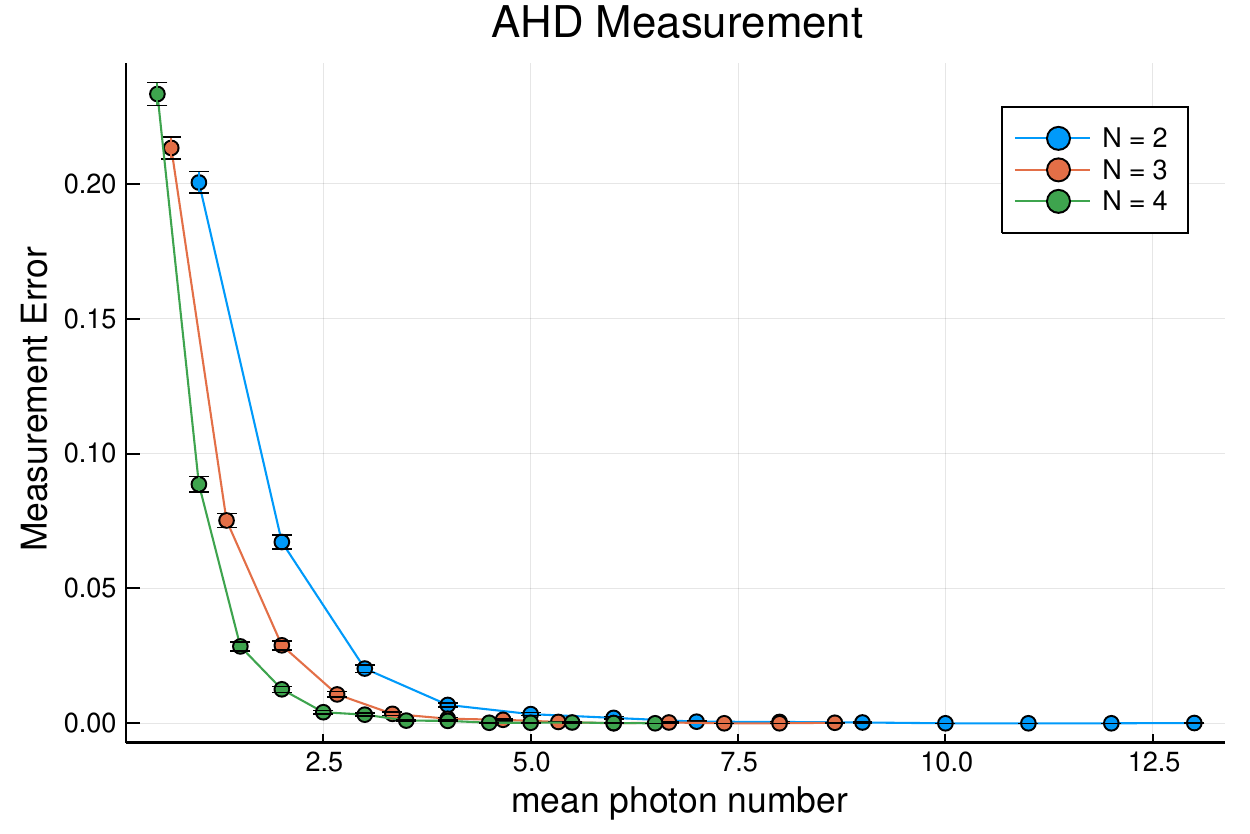}
\caption{Measurement Error vs $\bar{n}=1/2\cdot KN$ for binomial code with different $N$ values using AHD measurement.}
\label{fig:ahdmsmt}
\end{figure}

In Figs. \ref{fig:hetmsmterror},\ref{fig:ahdmsmt} the data for the measurement error rate as a function of mean photon number is generated numerically. We assume $\gamma = 0$ in order to decouple the measurement error from the noise due to photon loss. We then simulate many shots of qubit measurement and calculate the fraction of shots where the logical state as given by the Qubit State Inference, see Sec. \ref{singlequbitdecoding}, differs from the actual answer found according to Eq. \ref{sample}. By varying $K$, for fixed $N,$ we vary the mean photon number.

Fig. \ref{fig:hetmsmterror} shows the measurement error $q$ as a function of $\bar{n}$ for heterodyne measurement, for binomial codes of different $N.$ We see that for a given $q$, codes of higher $N$ will have a higher mean photon number. 

Fig. \ref{fig:ahdmsmt} shows measurement error as a function of mean photon number for AHD measurement. Not only is the overall measurement error much lower than in the heterodyne case, as expected, but also the measurement error is more consistent across $N$ values. As we will see in Sec. \ref{mainresults}, this has a significant impact on threshold values.

It is important to note that the `measurement error' we are referring to is measurement error at the level of the bosonic mode.  This manifests as Pauli errors of the binomial code qubits, rather than as errors in the binomial code single qubit $X$ measurement outcomes. Correcting binomial code qubit measurement outcomes would require multiple measurement rounds. Measurement error in our context contributes to Pauli noise independently from noise due to photon loss.

\subsection{Qubit State Inference}\label{singlequbitdecoding}
Once we have obtained the phase of an RSB encoded qubit, we must infer how the phase of the bosonic mode maps onto a $\pm1$ X basis qubit measurement outcome. In this section, we will quantify the performance of such \emph{qubit state inference} (QSI) techniques in terms of the fidelity of teleportation along a 3-qubit 1D cluster state as described in Sec.\ref{1Dcluster} using heterodyne measurement for the phase measurement.

\subsubsection{Binning Algorithm}
\begin{figure}
\begin{center}
\includegraphics[trim=4cm 14cm 4cm 4cm,clip,scale=0.4]{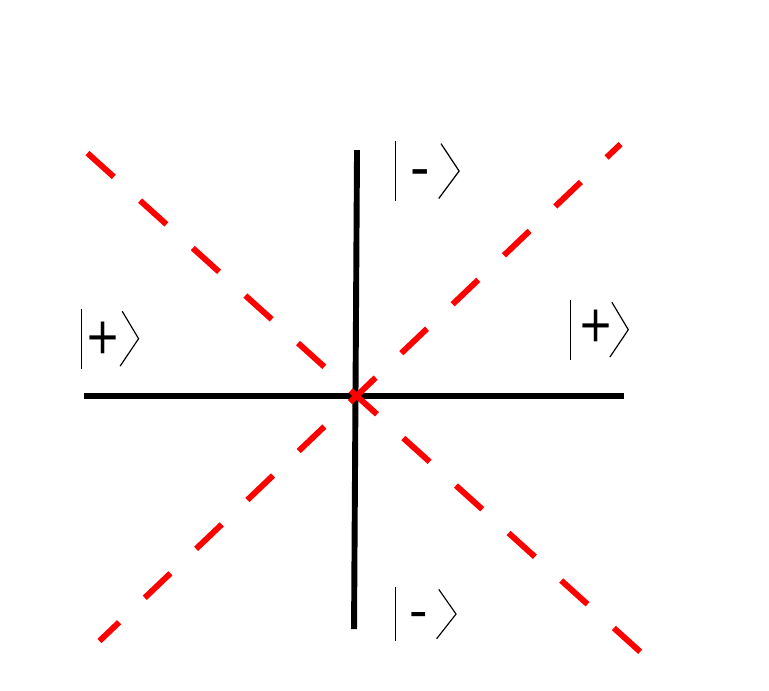}\\
\caption{Plot of binning regions in the complex plane for an N = 2 code.}
\label{fig:binning}
\end{center}
\end{figure}

The binning algorithm utilises the discrete rotational symmetry of RSB codes to straightforwardly bin the phase measurement in the complex plane according to the position of the logical plus and minus states in phase space. Fig. \ref{fig:binning} below represents a binning algorithm QSI for an $N = 2$ RSB code.

\subsubsection{Maximum Likelihood QSI}
For a three qubit 1D cluster state, the maximum likelihood QSI uses the phase measurement outcomes of the first two qubits to simultaneously infer the states of these qubits. It takes the full noise channel into account for each qubit, and therefore is the most accurate of the QSI techniques we examine. The maximum likelihood QSI works by calculating the probability of the two qubits being in states $i,j$ given their phase measurement outcomes $\phi_{1},\phi_{2}$ and choosing $i,j$ which maximise this probability. 

To design our QSI techniques we analyse their performance relative to a simplified noise model where the photon loss occurs before the $\textsc{CROT}$ gates. The loss may be commuted past the $\textsc{CROT}$ gates and will spread to dephasing noise on adjacent qubits. This is in contrast to how we model the noise to which the bosonic modes are subject in our full simulations where loss occurs $during$ the $\textsc{CROT}$ gates, as described in Sec. \ref{noise}. We use a simpler noise model to design our QSI techniques in order to make them tractable to implement and analyse numerically. Consequently, for the QSI techniques we model photon loss using Kraus operators for photon loss of order $k$
\begin{align}
    \hat{A}_{k}=\frac{1}{\sqrt{k!}}(1-e^{-\gamma})^{k/2}e^{\frac{-\gamma\hat{n}}{2}}\hat{a}^{k}.
\end{align}
Here the parameter $\gamma$ describes the loss probability for each photon. It corresponds to $\kappa t_{\rm gate}$ in the full model discussed previously in which photon losses can occur an any time.
The dephasing operators acting on qubit $i$, given by the commutator with the $\textsc{CROT}_{ij}$ of loss on qubit $j$, are
\begin{align}
    \hat{C}_{k} = e^{\frac{-i\pi k\hat{n}_i}{N^{2}}}.
\end{align}
Suppose that the first two qubits in the cluster have measurement outcomes $\phi_{1},\phi_{2}$ and we aim to infer qubit states $i_{1},i_{2}\in\{\pm1\}$ in order to obtain the Pauli correction we need to apply to qubit three. We will define the likelihood function $\mathcal{L}(i_{1},i_{2}|\phi_{1},\phi_{2})$ which is given by the conditional probability $p(\phi_{1},\phi_{2}|i_{1},i_{2})$ for the phase measurement outcomes given the initial qubit states. 

\begin{widetext} 
Each qubit of the cluster is subject to noise due to photon loss as well as dephasing from the photon loss ocurring on neighbouring qubits. Therefore
\begin{align} \label{maxlikelihood}
     \mathcal{L}(i_{1},i_{2}|\phi_{1},\phi_{2}) =  \sum_{k,k'=0}^{\infty}{\rm Tr}\left[\left(F(\phi_{1})\otimes \hat{F}(\phi_2)\right)\left(\hat{C}_{k'}\hat{A}_{k}|i_{1}\rangle\langle i_{1}|\hat{A}^{\dagger}_{k}\hat{C}^{\dagger}_{k'}\otimes \hat{C}_{k}\hat{A}_{k'}|i_{2}\rangle\langle i_{2}|\hat{A}_{k'}^{\dagger}\hat{C}^{\dagger}_{k}\right)\right].
\end{align}
Explicitly, maximum likelihood QSI chooses $i,j$ giving
\begin{align}\mathrm{argmax}_{i,j}\mathcal{L}(i,j|\phi_{1},\phi_{2}).
\end{align} \end{widetext}

For schemes with more qubits the exact maximum likelihood QSI can be formulated in the same way. However it becomes intractable as the number of qubits in the cluster state is increased. For this reason we consider an approximation that can be performed efficiently.

\subsubsection{Local Maximum Likelihood QSI}
We define a local maximum likelihood QSI that uses a local approximation of the noise channel acting on any given qubit to obtain an efficient approximation for the likelihood function. 

The form of the $\mathcal{L}(i_a|\phi)$, likelihood function for the qubit $a$ being in state $i$ conditioned upon its phase measurement outcome $\phi$, depends upon the number of qubits to which the qubit in question is entangled. 
Suppose that the qubits entangled to a specific qubit $a$ are denoted by $l\in\mathcal{N}(a).$ Let the state of qubit $l$ be $i_{l}$ and the state of \emph{all} the qubits will be indicated by $\vec{i}$. Let $\mathcal{S}$ denote the set containing all vectors $\vec{j}$ of length $|\mathcal{N}(a)|$ with entries $0\leq j_{l}\leq u$ indicating the number of photon emissions on the mode $l$. Let 
\begin{align}
    {\rho_{i_a}} &= \sum_{k=0}^{u}\hat{A}_{k}|i_a\rangle\langle i_a|\hat{A}_{k}^{\dagger}
\end{align} 
and
\begin{align}
\mathcal{O}_{a,\vec{i}}&=\sum_{\vec{j}\in\mathcal{S}}p_{\vec{i}}(\vec{j})\hat{C}_{\vec{j}}\rho_{i_a}\hat{C}^{\dagger}_{\vec{j}}
\end{align} Where the \emph{a priori} photon emission probabilities are
\begin{align}
    p_{\vec{i}}(\vec{j})&=\prod_{l\in\mathcal{N}(a)}p_{i_{l}}(j_{l})\nonumber \\
    &=\prod_{l\in\mathcal{N}(a)}\mathrm{Tr}(\hat{A}_{j_{l}}|i_{l}\rangle\langle i_{l}|\hat{A}_{j_{l}}^{\dagger})
\end{align} and \begin{equation}
    \hat{C}_{\vec{j}}=\prod_{l\in \mathcal{N}(a)}\hat{C}_{j_l}
\end{equation}

Then we define the approximate likelihood function to be\begin{align}\label{localml}
    \mathcal{L}_{LL}(\vec{i}|\phi)&=\prod_{a=1}^M
    \mathrm{Tr}\left[\hat{F}(\phi)\mathcal{O}_{a,\vec{i}}\right]
\end{align}
%\begin{align}
%    p(i|\phi) &=\sum_k\sum_{k'_1,\ldots k'_{|\mathcal{N}(a)|}}\sum_{j_1,\ldots,j_{|\mathcal{N}(a)|}}\mathrm{Tr}\left[\hat{F}(\phi)\hat{D}_{\mathbf{k}'}\hat{A}_{k}|i\rangle\langle i|\hat{A}^{\dagger}_{k}\hat{D}_{\mathbf{k}'}^{\dagger}\right]
%\end{align}
where we truncate the two sums over numbers of photon losses at $u$ for numerical tractability. Given $p(i|\phi)$ the local maximum likelihood QSI works by choosing $i$ as follows
\begin{align}
\mathrm{argmax}_{\vec{i}}\mathcal{L}_{LL}(\vec{i}|\phi).
\end{align}

\subsection{1D Performance Metric}
Having used a QSI procedure to determine the states of qubits in the chain up to the final qubit $M$, we apply a Pauli operation that depends on these outcomes. In the ideal case by applying the recovery operator to the final qubit, we obtain the original logical state on that qubit. We can write the effect of both the QSI procedure and the recovery operation as a CPTP map $\mathcal{R}(\vec{\phi})$ where $\vec{\phi}$ is the set of all $M-1$ measurement outcomes.

%\begin{widetext}
The quantum channel representing the overall teleportation operation, including encoding into a linear cluster state chain, photon loss, phase measurement QSI and recovery is 
\begin{align}
    &\mathcal{E}(|\psi\rangle \langle \psi|)=\int d\vec{\phi}\mathcal{R}(\vec{\phi})\mathrm{Tr}_{1\ldots M-1}\left[F(\vec{\phi}) \tilde{\Gamma}(\rho_{\psi})\right]\mathcal{R}^{\dagger}(\vec{\phi})
\end{align}%\end{widetext}
where
\begin{align}
    \rho_\psi = U\left(|\psi\rangle\langle\psi|\otimes|+\rangle\langle+|^{\otimes M-1}\right)U^\dagger\nonumber
\end{align}
and \begin{equation*}
    U=\prod_{i=1}^{M-1}\textsc{CROT}_{i,i+1}.
\end{equation*}
The quantum channel $\mathcal{E}$ acts on an initial qubit state $|\psi\rangle$ and maps it to an output qubit state encoded in the final RSB qubit of the cluster.
$\tilde{\Gamma}$ is the noise channel commuted past the $\textsc{CROT}$ gates as defined as in Eq. \ref{eq10} and   $F(\vec{\phi}) = \otimes_{i=1}^{M-1}F(\phi_{i})$ are the measurement operators..

We will use the entanglement fidelity as the performance metric to quantify the performance of the 1D cluster state.
We follow the standard definition of entanglement fidelity \cite{nielsenchuang} of the quantum channel by supposing that our system of interest $Q$, in our case the final qubit of the cluster state, is coupled to an ancilla qubit $R$ in a reservoir and letting
\begin{align}
    F(\rho,\mathcal{E}) &= \langle RQ|(\mathcal{I}_{R}\otimes\mathcal{E})(|RQ\rangle\langle RQ|)|RQ\rangle
\end{align}
where $\mathcal{I}_{R}$ denotes the identity channel on the ancilla qubit. $|RQ\rangle$ may be any maximally entangled state but we choose $|RQ\rangle = (|00\rangle+|11\rangle)/\sqrt{2}$.

\subsection{1D Cluster State Results}\label{1dresults}

\subsubsection{Numerical Performance of Phase Measurements}

\begin{figure}[h]
\includegraphics[scale=0.4]{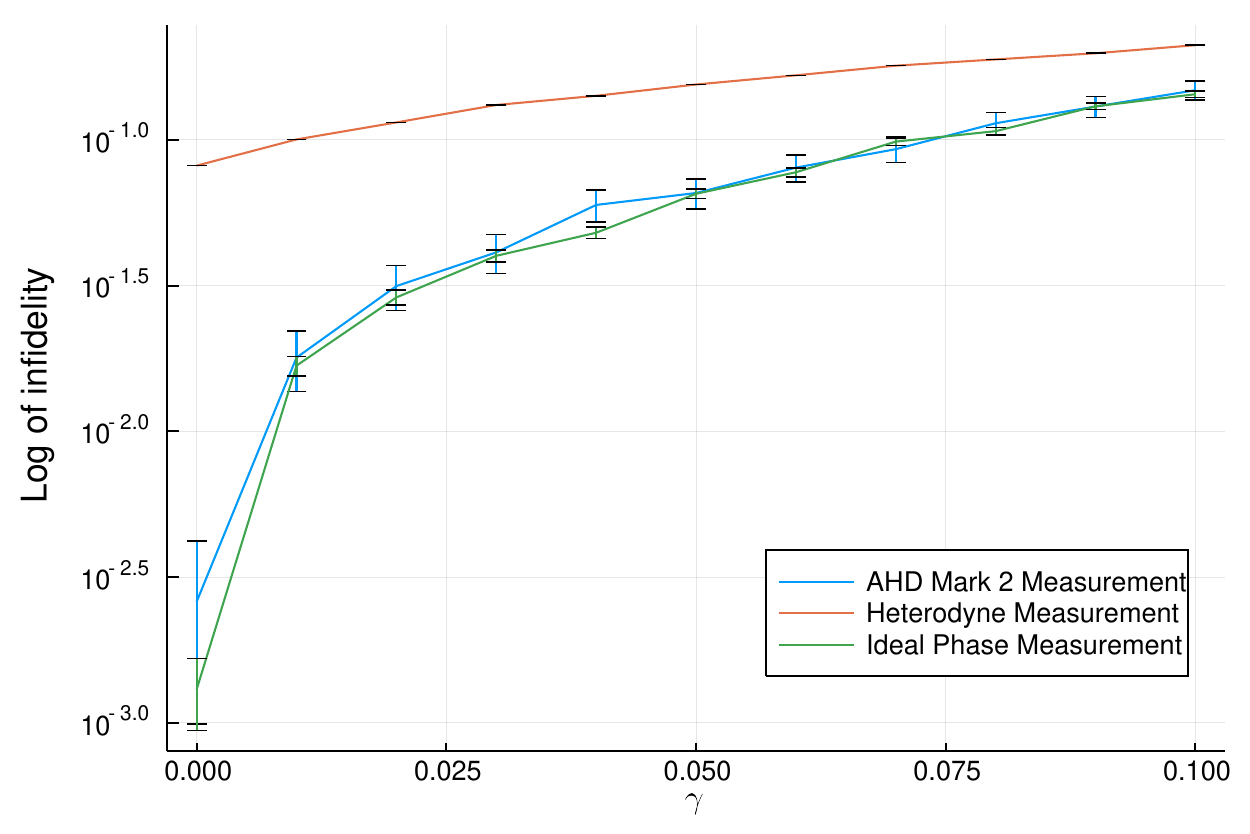}
\caption{Comparison of measurement schemes using log of Infidelity vs $\gamma$, strength of photon loss, for a 3-qubit cluster state with binomial encoded qubits, $N = 2, K = 6$.}
\label{fig:msmtperformance}
\end{figure}

We performed simulations of the minimal instance of the binomial encoded 1D cluster state scheme with a 3-qubit cluster state. 
Fig. \ref{fig:msmtperformance} below quantifies the performance of the different measurement schemes. We see that both canonical phase and AHD measurement schemes significantly outperform heterodyne. Furthermore, there does not appear to be a notable difference between the performance of AHD and canonical phase measurements. This indicates that AHD is a good alternative to the optimal canonical phase measurement. Note that the nonzero infidelity at zero $\gamma$ is due to the measurement error inherent described in Sec. \ref{msmterrorsection}. Even in the absence of photon loss, due to the finite mean photon number of the binomial encoded qubits, measurement error will lead to qubit-level errors.

\subsubsection{Qubit State Inference Comparison} \label{decoders}
We now examine a numerical comparison of the performance of the binning algorithm QSI, maximum likelihood QSI and local maximum likelihood QSI for the three qubit cluster state telecorrection scheme. These qubit state inference techniques are described in detail in Sec. \ref{singlequbitdecoding}.

\begin{figure}[h]
\includegraphics[scale=0.4]{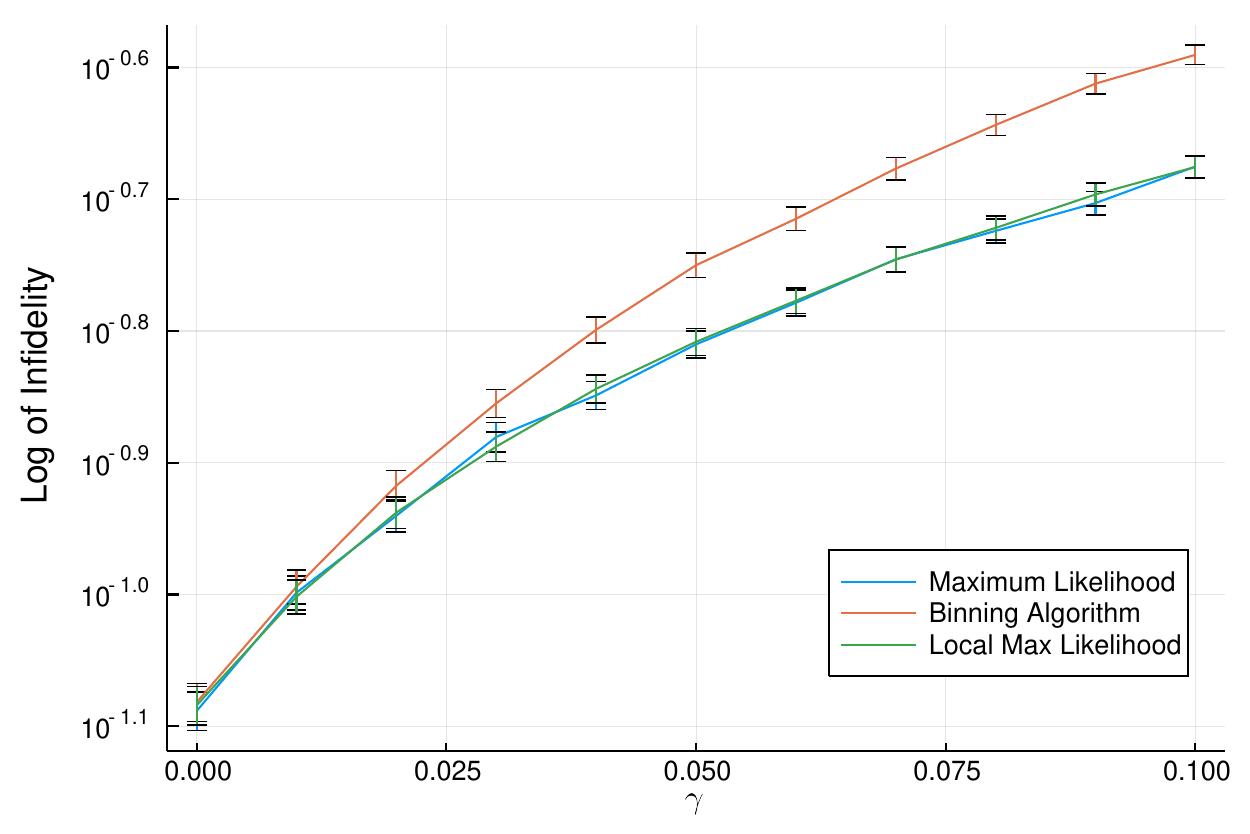}
\caption{Comparison of QSI techniques for a 3-qubit increases uster state with binomial code qubits with $N = 2, K = 6$.}
\label{fig:decodercomparison}
\end{figure}

Fig. \ref{fig:decodercomparison}  shows the performance of the three QSI techniques for a 3-qubit cluster state with binomial code qubits, as a function of the loss parameter $\gamma.$ As expected, for no loss the QSI technqiues converge on the same value of infidelity. However as $\gamma$ is increased, the binning algorithm performs significantly worse than both the maximum likelihood and local maximum likelihood QSI techniques. Notably, the performance of maximum likelihood and local maximum likelihood is statistically identitical here, indicating that the local maximum likelihood QSI is a good replacement for the full maximum likelihood QSI.

Together these results indicate that adaptive phase measurement combined with a local maximum likelihood QSI is a considerable improvement on the straightfoward heterodyne and binning approach to qubit measurement for the binomial code.

\subsection{2D Surface Code Methods}\label{2danalysis}

To enable simulations of the full 2D surface code we need to introduce some additional methods. In the following we will discuss the twirling methods that we used in order to have tractable simulations as well as the details of the surface code decoder that we implemented.

\subsubsection{X-Basis Pauli Twirl} \label{twirl}
For the purpose of numerical tractability when simulating these larger systems, we introduce an X-basis Pauli twirl for the RSB qubits. We show in Appendix~\ref{Xbasispaulitwirl} that in the limit where $\gamma=0$ the twirl has no effect on the measurement statistics and this approximation becomes exact. For non-zero photon loss the twirled noise model is only approximately the same as the physical photon loss model. By removing certain coherences in the photon loss error model the twirl enables simulations to find the threshold of the planar code. We do not expect that the twirl qualitatively changes our results. Simulations that avoid it would require much more expensive simulations, for example based on tensor network simulations~\cite{darmawan2017tensor}.

We define the twirl as a mapping $\mathcal{P}_{X_{i}}$ to each qubit $i$ of the RSB code cluster state. 
The X-basis Pauli twirl acting on a state $\rho$ is given by
\begin{align}
    \mathcal{P}_{X_{a}}(\rho) &= |+\rangle\langle+|_{a}\otimes\langle+|\rho |+\rangle_{a}+|-\rangle\langle-|_{a}\otimes\langle-|\rho |-\rangle_{a}.
\end{align}
Note that while the twirl defined here affects qubit $a$, the density matrix $\rho$ describes the state of the whole set of qubits, not all of which are affected by the twirl. The operator $\langle-|\rho |-\rangle_{a}$ acts on the qubits that are not affected by the twirl. 

Consider the first $X$ measurement to be performed on qubit $a$ given times and locations of the photon emissions labelled by $\mathbf{t}$. The state of the remaining qubits for a given phase measurement outcome $\phi_a$ will be \begin{equation}
\rho_{\phi_a,\mathbf{t}}=\mathrm{Tr}_a[\hat{F}_a(\phi_a)\tilde{\Gamma}_{\mathbf{t}}(\rho)]
\end{equation}
The norm of this state relates to the probability of the measurement outcome, and in fact we will only be interested in integrals over values of $\phi_a$ that map to a given outcome $\pm$ for qubit $a$.

In our simulations we replace these states with states where a $X$-basis twirl is applied prior to the measurement: \begin{equation}
\bar{\rho}_{\phi_a,\mathbf{t}}=\mathrm{Tr}_a\{\hat{F}_a(\phi_a)\tilde{\Gamma}_{\mathbf{t}}[\mathcal{P}_{X_a}(\rho)]\}
\end{equation}

To see why this replacement could be justified, let $M_{\pm}$ denote the POVM elements of the noise-plus-measurement-plus-QSI process that is applied to the ideal RSB code qubits and aggregates the phase measurements into the $\pm 1$ outcomes for the measurement. The application of the Pauli twirl $\mathcal{P}_{X}$ to this encoded ideal initial state is justified so long as $M_{\pm}$ is diagonal in the X-basis of the RSB code qubit since in this case $\mathcal{P}^\dagger_X[M_\pm]=M_\pm$. For the case that $\gamma=0$ we provide a proof that $\mathcal{P}^\dagger_X[M_\pm]=M_\pm$ in the case of the binning QSI in Appendix \ref{Xbasispaulitwirl}. 

Note that this replacement works only in the case where, as here, we are interested in some averaged quantity for the schemes such as average entanglement fidelity for a logical qubit channel. If we inspected the data conditioned on phase measurements analysed in some more fine grained way then this twirling approximation is not necessarily justified.

This twirl dramatically simplifies our numerical implementation. 
Suppose that the ideal initial cluster state of the scheme is $\rho$. So in particular $\rho$ is in the codespace of the RSB code.  Consider measuring qubit $a$ of the cluster as before.  Then measuring qubit $a$ of an $M$ qubit cluster state will result in the state
\begin{align} \label{pteq}\rho_{\phi_a,\mathbf{t}}=&\mathrm{Tr}_{a}\{\hat{F}_a(\phi_a)\tilde{\Gamma}_{\mathbf{t}}[\mathcal{P}_{X_a}(\rho)]\}\nonumber\\
    &=\langle +|\Gamma^\dagger_{a,\mathbf{t}}[\hat{F}_a(\phi_a)]|+\rangle\tilde{\Gamma}_{\mathbf{t}/a}[\langle+|\rho|+\rangle_a ]\nonumber \\ 
    &+ \langle -|\Gamma^\dagger_{a,\mathbf{t}}[\hat{F}_a(\phi_a)]|-\rangle\tilde{\Gamma}_{\mathbf{t}/a}[\langle-|\rho|-\rangle_a ]
\end{align}
This is a state on $M-1$ qubits. The adjoint noise map on qubit $a$ is \begin{align}
\tilde{\Gamma}_{a,\mathbf{t}}^{\dagger}(\cdot) &= \hat{E}_{a,\mathbf{t}}^{\dagger}\cdot \hat{E}_{a,\mathbf{t}}
\end{align}
where $\hat{E}_{a,\mathbf{t}}$ denotes the loss operator acting on qubit $a$ with the times and locations of photon emissions given by $\mathbf{t}$. The noise map on all qubits other than $a$ is \begin{align}
\tilde{\Gamma}_{\mathbf{t}/a}(\cdot) &= \left(\prod_{i\neq a}\hat{E}_{i,\mathbf{t}}\right)\cdot \left(\prod_{i\neq a}\hat{E}_{i,\mathbf{t}}\right)^{\dagger}
\end{align}

Eq.~\ref{pteq} is simply stating that after the the measurement, the remaining cluster state is in a mixture of ``qubit $a$ was in the logical plus state when measured", and ``qubit $a$ was in the logical minus state when measured", with probabilities
\begin{align}
    p(\pm|\phi_{a},\mathbf{t}) &= \frac{\langle \pm |\Gamma^\dagger_{a,\mathbf{t}}[\hat{F}_a(\phi_a)]|\pm \rangle}{\mathrm{tr}\{\tilde{\Gamma}^{\dagger}_{\mathbf{t}}[\hat{F}_a(\phi_{a})\tilde{\rho}]\}} \label{sample}
\end{align}
where $\tilde{\rho} =\mathcal{P}_{X_{a}}(\rho),$ and the denominator of the above expression can be inferred by the normalisation requirement. Note that to obtain this we have used the fact that $\mathrm{Tr}\langle +|\rho|+\rangle_a=\mathrm{Tr}\langle -|\rho|-\rangle_a$ which is a readily verified property of the ideal cluster state $\rho$. While we have focussed here on a single measurement, clearly each successive measurement can be treated in the same way since after the first measurement the states $\langle \pm|\rho|\pm\rangle_a$ are again ideal cluster states in the RSB code subspace.

In our numerical simulations we sample using these probabilities to determine which state $|\pm\rangle\langle\pm|_{a}$ the qubit was in  when measured. The Pauli X-basis twirl is implicit in this step of the algorithm. We then compare this outcome to the outcome we get by inferring the qubit state from the phase measurement outcome $\phi$. A bit flip error is placed on the qubit if the `actual' state of the qubit disagrees with the inferred measurement outcome.
 
\subsubsection{Modified MWPM Decoder}\label{dijk}
We can further optimise the performance of RSB codes in our concatenated model by modifying the standard MWPM decoder. The standard MWPM decoder works by taking the error syndromes and forming a complete graph, where edge weights between nodes are given by the Manhattan distance - that is, each each edge of the lattice is assigned unit weight. We can use a modified MWPM decoder to assign edge weights based on phase measurement results. We follow the approach of \cite{softmeasurement} by utilising the information contained in realistic phase measurements which is discarded when mapping to a binary X basis measurement outcome. As in \cite{softmeasurement}, we adopt the concepts of hard and soft measurement. In our model, the phase measurement of the RSB qubit gives the `soft' measurement outcome $\phi$, which is then processed by a QSI technique and mapped onto an observed `hard' outcome $\hat{\mu}\in\{+1,-1\}$, corresponding to the $|\pm\rangle$, respectively. Recall the discussion in Sec. \ref{twirl} in which we stated that after subjecting a qubit of the cluster to the X-basis Pauli twirl, we sample probabilistically according to Eq.~\ref{sample} to determine whether the qubit was in the state $|\pm\rangle\langle\pm|_{i}$ when measured. We define the `ideal' hard outcome $\bar{\mu}\in\{\pm 1\}$ to be this sampled outcome, which is kept track of in our numerical simulation but is not accessible to the decoder.

During the MWPM algorithm, a complete graph is created from nodes of the syndrome graph. Distances are then calculated between all possible combinations of node pairs. If, for a given node pair, edge weights of the syndrome graph are calculated using soft measurement outcomes, distances cannot be precomputed and must be evaluated using modified MWPM algorithm. This means that for each node pair in the complete graph, modified MWPM algorithm must be run again. In contrast, if edge weights for a node pair are given fixed unit weight, modified MWPM algorithm need only be run a single time. We use a hybrid model in which for nodes separated by Manhattan distances greater than $\log_{2}(L)$, edge weights are calculated using observed hard measurement outcomes, whilst for nodes pairs for which $d_{Manhattan} \leq \log_{2}(L)$, edge weights are calculated using soft measurement outcomes. This provides the advantage of exploiting soft information for a more accurate syndrome decoder, whilst avoiding the long runtime of exhaustively calculating shortest paths between every possible node pair combination.

Let the soft outcome observed be $\phi$, the probability distribution function of which, conditioned on the ideal hard outcome $\bar{\mu}=\pm 1$, is $f^{\bar{\mu}}(\phi)$. For the maximum likelihood and local maximum likelihood QSI techniques, $f^{\bar{\mu}}(\phi)$ is given by Eqs. \ref{maxlikelihood} and \ref{localml} respectively. The soft outcome is then mapped onto the observed hard outcome $\hat{\mu}$ according to to \textit{hardening map}
\begin{align}
    \hat{\mu} = 
    \begin{cases}
    &+1,f^{+1}(\phi)>f^{-1}(\phi)\\
    &-1, f^{-1}(\phi) > f^{+1}(\phi).
    \end{cases}
\end{align}
The QSI technique we choose gives the explicit form of the hardening map. In our 2D surface code simulations, the two QSI techniques we employ are the binning algorithm QSI and local maximum likelihood QSI. The full maximum likelihood QSI would be too computationally expensive.

We say that a bit flip occurs on the qubit if the observed hard outcome disagrees with the ideal hard outcome. Again, this information is accessible only to the simulation and not to the decoder.

As mentioned, a standard MWPM decoder proceeds by assigning each edge of the surface code lattice unit weight, $w_{e} = 1.$ This is equivalent to saying that each qubit is equally likely to have experienced an error. However this assumption is not accurate in general and the information contained in the soft measurement outcomes can be utilised to weight the qubits (edges) according to the probability that the observed hard outcome disagrees with the ideal hard outcome, and thus the probability that the qubit experienced an error. We define the likelihood ratio for a given qubit
\begin{align}
    L(\phi) &= \frac{f^{\hat{\mu}'}(\phi)}{f^{\hat{\mu}}(\phi)}
\end{align}
where $\phi$ is the soft measurement outcome, $\hat{\mu}$ is the observed hard measurement outcome and $\hat{\mu}'$ is the `other' value of the hard outcome, ie $\hat{\mu} = +1 \implies \hat{\mu}' = -1.$ Edges for which the soft measurement outcome is used to determine the weight are then weighted according to
\begin{align}
    w &= -\log[L(\phi)]
\end{align}
whilst edges whose weights are determined by the observed hard measurement outcome are weighted as
\begin{align}
    w = -\log[p_{e}/(1-p_{e})]
\end{align}
where $p_{e}$ is the physical error rate. In our model, the physical error rate is given by the effective bit flip rate due to measurement error and photon loss, and can be calculated numerically. This is done by, for a given loss rate, simulating the code subject to loss and averaging over the number of qubits suffering a bit flip to obtain an effective bit flip rate.

\subsubsection{2D Performance Metrics}
We quantify the behaviour of the 2D foliated surface code in terms of its threshold value.

The threshold of a code refers to the error rate below which increasing the size of the code decreases the logical error rate. For realistic computations, it will be necessary to operate in the sub-threshold regime. Therefore, having a high threshold is a desirable property as it reflects a code's capacity to tolerate noise whilst still being able to do computations. 

Finally, we note that we benchmark the performance of the binomial code against the trivial Fock space encoding, as well as binomial codes of different orders of discrete rotational symmetry. 

\subsubsection{Overview of Numerical Model}
We now provide a sketch of the numerical model used.

    \textit{Implementation of noise:} As described in Eq. \ref{noise} and in Appendix\ref{quanttraj} the photon emission times for each qubit can be sampled independently according to a probability distribution that ignores the $\textsc{CROT}$ gates. The outcome of this sample is an array $\mathbf{t}$ describing the times and locations of photon emissions. Given this the error operator for each qubit $\hat{E}_{a,\mathbf{t}}$ is determined and the overall error operation is just the product of each of these single-qubit error operators.

   \textit{Phase measurement:} Iterating over the lattice, each qubit is subject to a phase measurement modelled by a rejection sampling algorithm with $\phi\in[0,2\pi]$. Assuming a phase measurement with POVM $\hat{F}(\phi),$ the probability distribution used in rejection sampling is $p_a(\phi) = \mathrm{tr}[\hat{F}_a(\phi)(\hat{E}_{a,\textbf{t}}\rho \hat{E}_{a,\textbf{t}}^{\dagger})]$ where the error operator for the qubit $a$ being measured is $\hat{E}_{a,\textbf{t}}$ as defined in Eq. \ref{eq10}, and $\rho = \frac{1}{2}(|+\rangle\langle+|+|-\rangle\langle-|)$ is the maximally mixed state. The initial state of the qubit is taken as the maximally mixed state which is the reduced density matrix of any single qubit in a cluster state, as can be shown by  a simple stabilizer argument. This probability distribution corresponds to the $X$-basis twirled phase measurement probability distribution implied by Eq.  ~\ref{pteq}. If the qubit being measured is primal, the measurement outcome is stored in a primal measurement outcome array, and similarly if the qubit is dual.
   
    \textit{Single-qubit decoding:} The binning algorithm QSI requires no input other than the soft measurement outcome of the qubit in question, and uses the discrete rotational symmetry of RSB codes to bin the measurement outcome in the complex plane, as previously described. The local maximum likelihood QSI uses the (soft) measurement outcome of each qubit, as well as the (hard) inferred measured outcomes of neighbouring qubits to determine the explicit form of the probability distribution $p(\phi|i=\pm1)$ to be maximised. Soft measurement outcomes are inferred sequentially across the lattice left to right, top to bottom, so that each qubit has already had measurement outcomes of neighbours `above' it in the lattice inferred to use as input for its own QSI. Once the soft measurement outcome of the qubit has been inferred, for dual qubits we are done. For primal qubits we determine the `actual state' of the qubit as described in Sec. \ref{twirl}. The `actual state' of the primal qubit is compared to the inferred measurement outcome. If they disagree, a bit flip error is placed on that qubit which is represented by a $-1$. Otherwise, no error has occurred and the qubit is assigned the value $+1.$
    
    \textit{Error correction:} We begin by measuring logical Z before error correction. We choose one of the smooth boundaries. If an odd number of boundary qubits have errors, the logical Z measurement is $-1$, otherwise it is $+1$. Error correction proceeds as in Sec. \ref{mbqc}. Stabiliser measurements are modelled by, for each plaquette, multiplying together the stored values of all primal qubits comprising the plaquette. If a plaquette has an odd number of qubits with errors, there will be an odd number of $-1$ values and so the result of the multiplication will be $-1$. Otherwise, it will be $+1$. Stabilisers with $-1$ measurement outcomes form the error syndrome. The BlossomV implementation of the MWPM algorithm is then used to pair up nodes in the error syndrome. As we are modelling the planar code, there can be an odd number of $-1$ stabiliser outcomes, so it is also possible to pair syndrome nodes with the boundary. If the modified MWPM algorithm decoder is being used, it is used as described in Sec. \ref{dijk} to calculate weights between pairs of syndrome nodes to feed into MWPM. Otherwise, each edge of the lattice is assigned weight one. The now paired syndrome nodes are used to determine if a logical error has occurred. We pick the same smooth boundary we used for our initial logical Z measurement. For each of the paired syndrome nodes, we check if the path between them crosses our chosen smooth boundary. If an odd number of crosses occur out of all pairs, the logical Z measurement is $-1$, otherwise it is $+1.$ We compare the logical Z measurement before correction to the new logical Z measurement. If they disagree, a logical error has occurred.

\subsection{2D Surface Code Results}\label{mainresults}

In this section we will discuss the results of our simulations of the planar code concatenated with binomial codes under loss errors. 
\subsubsection{Threshold curves}
\begin{figure}[h]
\includegraphics[scale=0.4]{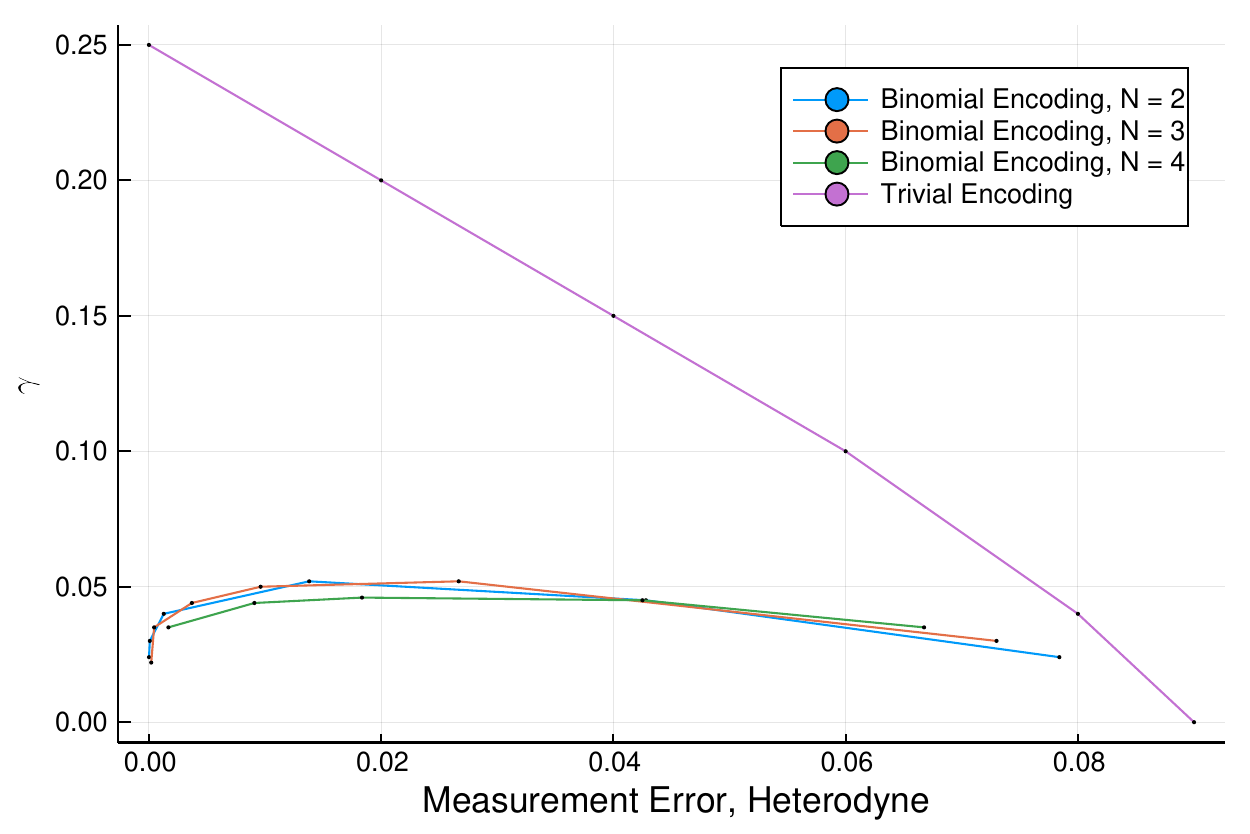}\\
\caption{Threshold curve for loss errors $\gamma=\kappa t_{\rm gate}$ for heterodyne phase measurement and binning QSI (Method I).  The trivial encoding with measurement error probability $q$ and amplitude damping noise $\gamma$ is compared against the binomial encodings, for different values of $N$ and $K$. The threshold $\gamma$ for each value of $N$ and $K$ is plotted at the corresponding effective qubit measurement error rate. There is no apparent advantage to the binomial code, and each value of $N$ and $K$ has roughly the same performance.}
\label{fig:basicthresh}
\end{figure}
\begin{figure}[h]
\includegraphics[scale=0.4]{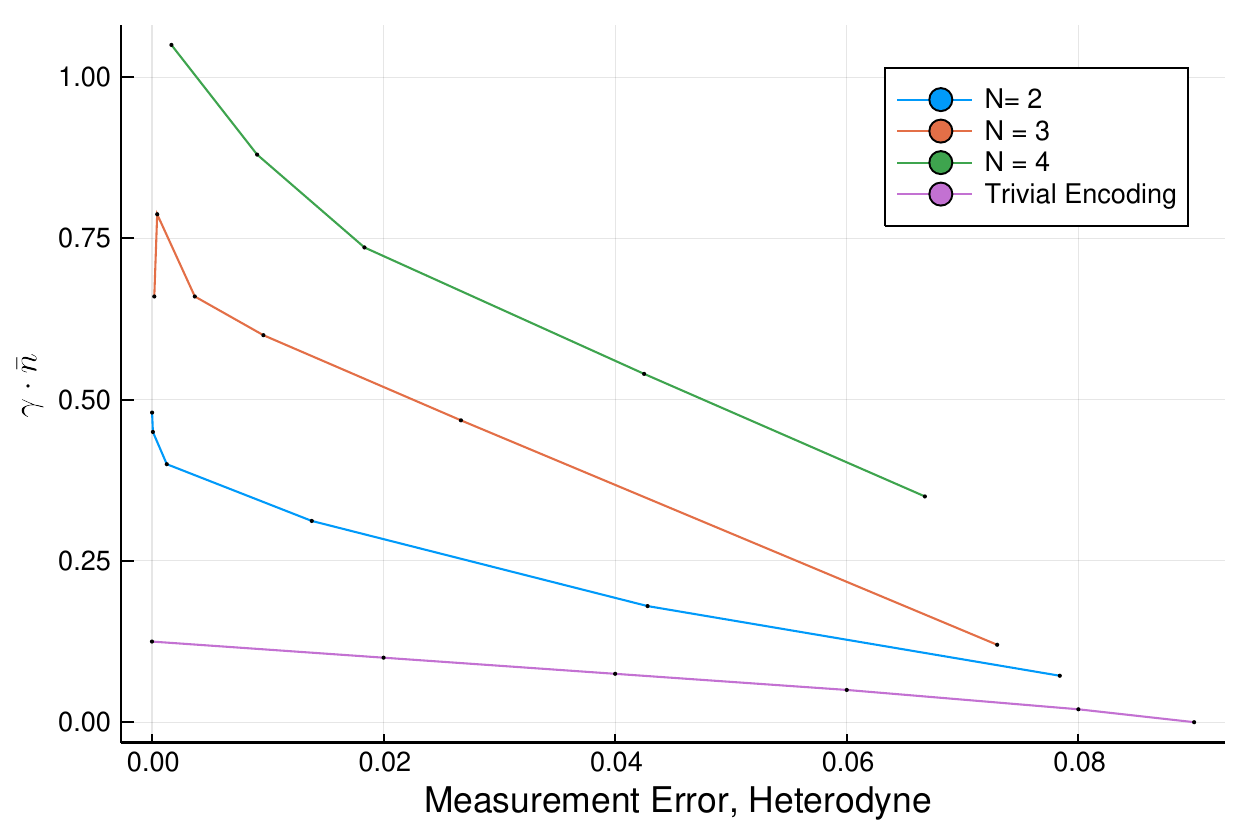}\\
\caption{The thresholds from Fig.~\ref{fig:basicthresh} are now plotted against the quantity $\bar{n}\cdot\gamma,q$. The probability of a single photon loss scales as $\gamma\cdot\bar{n}$, showing that higher $N$ codes do indeed correct more loss errors. Trivial encoding is compared against the binomial encoding, for different values of $N$ and $K$ and by this measure the trivial encoding is inferior to the binomial codes. }
\label{fig:normalisedthresh}
\end{figure}

\begin{figure}[h]
\includegraphics[scale=0.4]{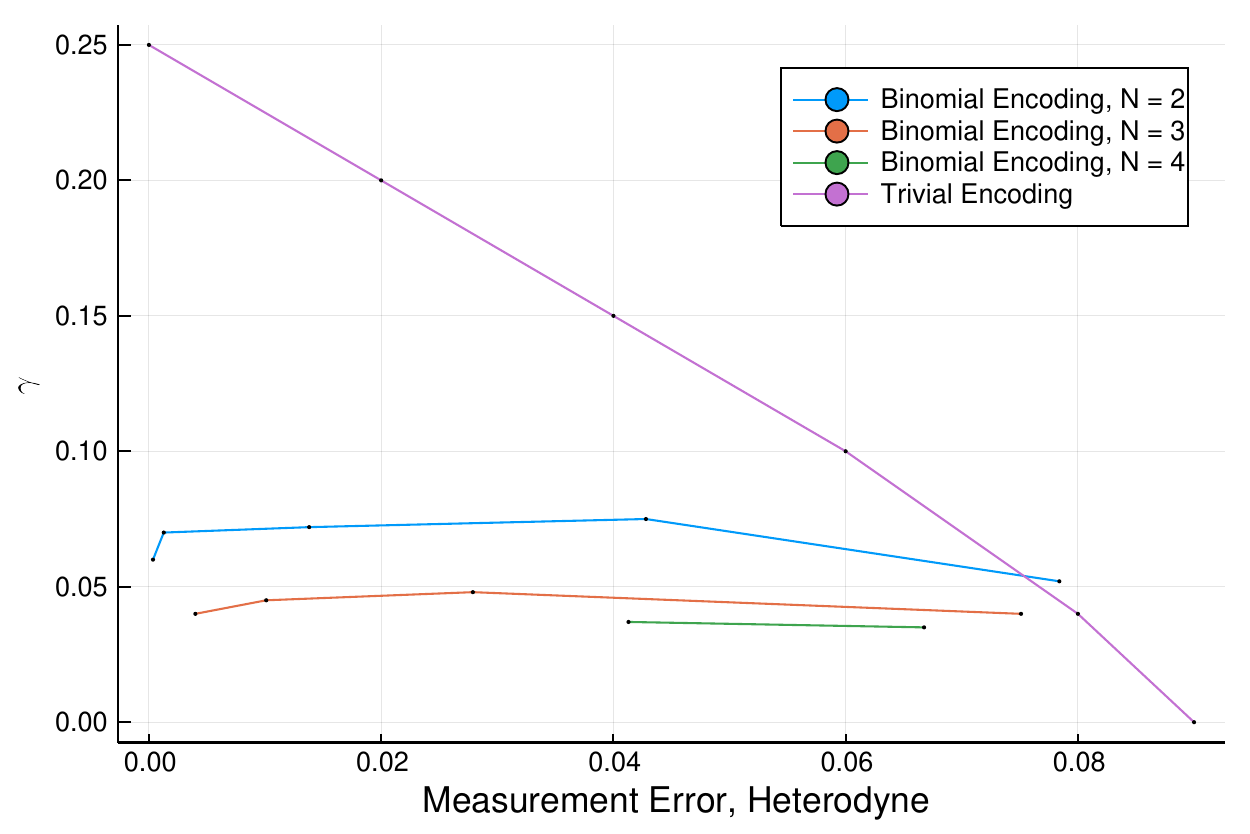}\\
\caption{Threshold for photon loss errors $\gamma$ with improved quantum state inference based on heterodyne phase measurement. The trivial encoding for photon loss $\gamma$ and measurement error $q$ is compared against the binomial encoding for different values of $N$ and $K$ with each data point being plotted at the effective qubit measurement error rate for that value of $N$ and $K$.  Local maximum likelihood quantum state inference is used to process heterodyne phase measurements, showing a performance improvement from switching QSI techniques.}
\label{fig:hetlocalml}
\end{figure}

\begin{figure}[h]
\includegraphics[scale=0.4]{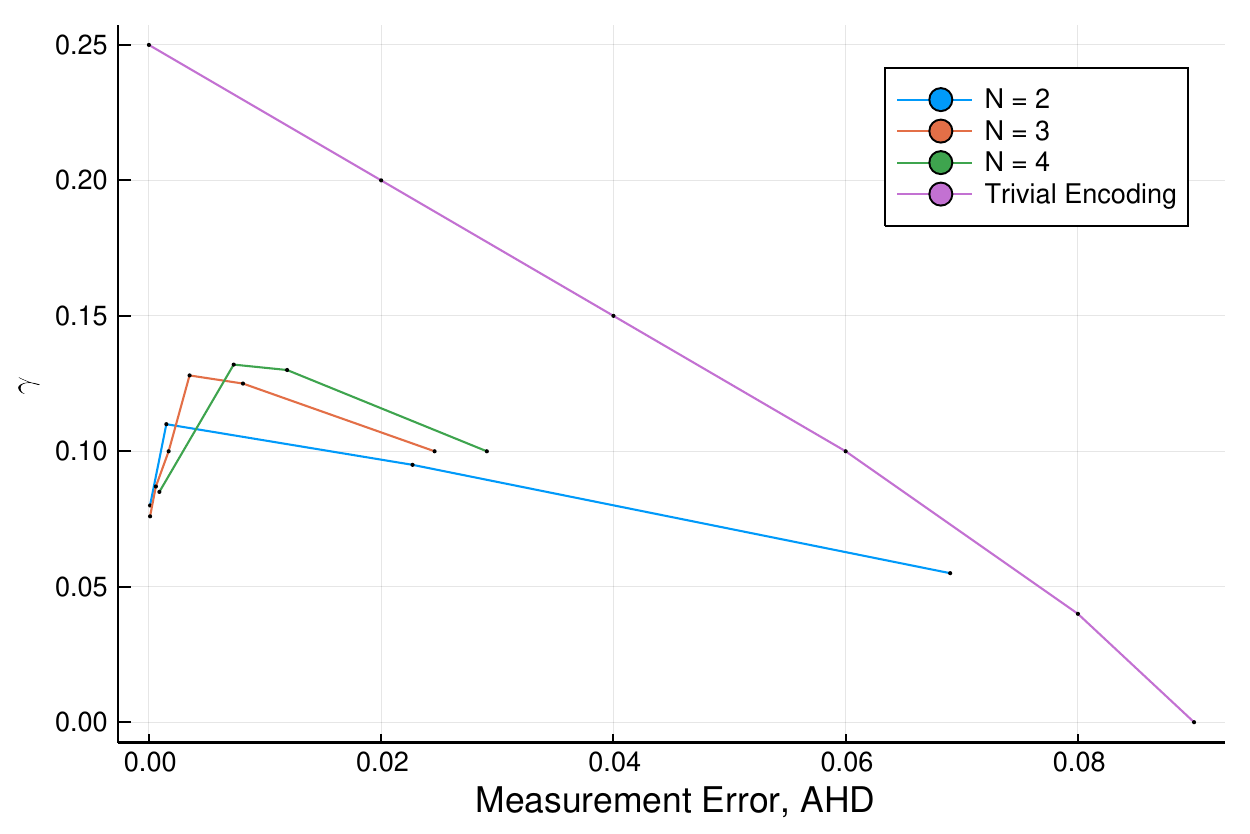}\\
\caption{Threshold for photon loss errors $\gamma$ with improved phase measurements as well as quantum state inference.  The trivial encoding for photon loss $\gamma$ and measurement error $q$ is compared against the binomial encoding for different values of $N$ and $K$ with each data point being plotted at the effective qubit measurement error rate for that value of $N$ and $K$. Method 2, which involves both local ML QSI and adaptive phase measurement, is used. Again, we see a performance improvement from changing phase measurement methods.}
\label{fig:ahdlocalml}
\end{figure}

\begin{figure}[h]
\includegraphics[scale=0.4]{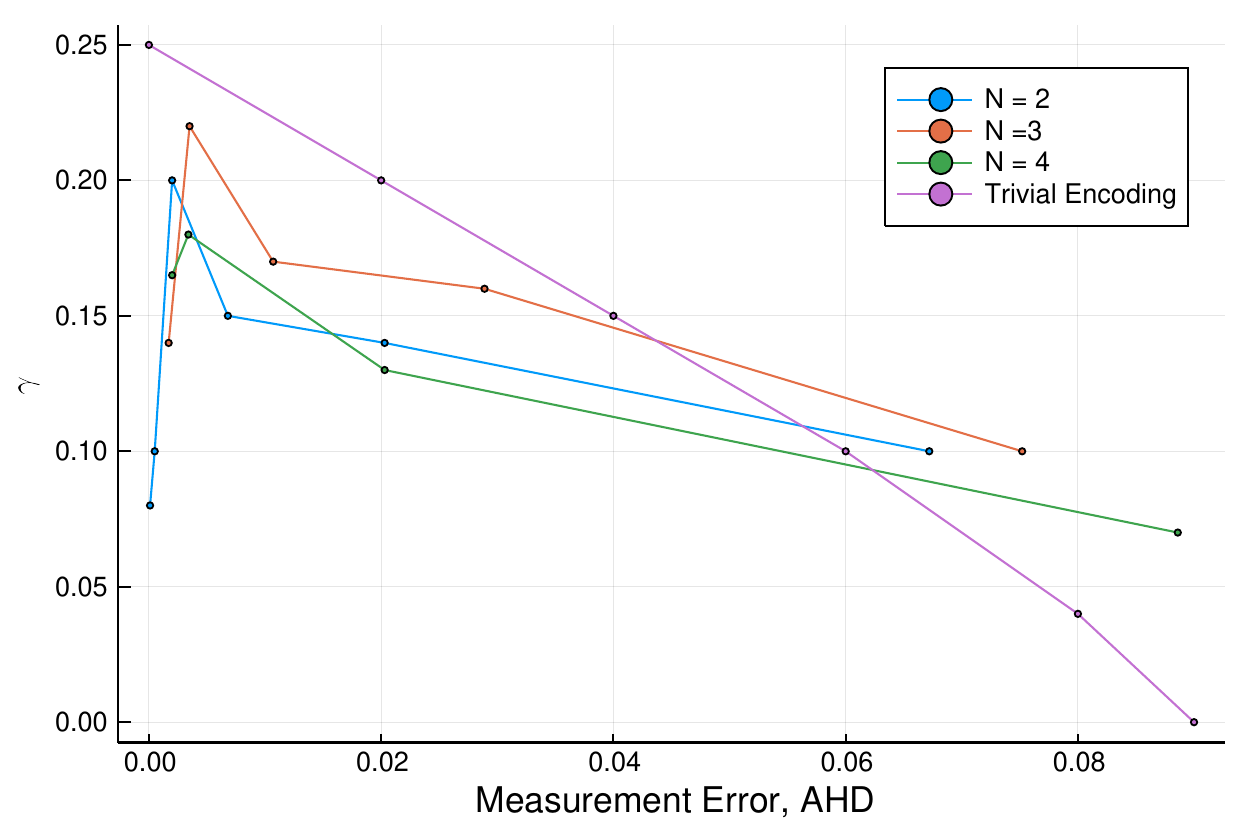}\\
\caption{Threshold for photon loss errors $\gamma$ with modified MWPM decoding.  The trivial encoding for photon loss $\gamma$ and measurement error $q$ is compared against the binomial encoding for different values of $N$ and $K$ with each data point being plotted at the effective qubit measurement error rate for that value of $N$ and $K$. Method 2, which involves both local ML QSI and adaptive phase measurement, is used. Thresholds of the binomial codes are now competitive with the trivial encoding at comparable measure error levels.}
\label{fig:dijkstraphase}
\end{figure}

In this section we will refer to codes using heterodyne phase measurement and the binning algorithm QSI as Method 1 Codes, and codes using AHD measurement and local ML QSI as Method 2 Codes. Figs. \ref{fig:basicthresh},\ref{fig:hetlocalml},\ref{fig:ahdlocalml},\ref{fig:dijkstraphase} show the threshold for different codes as a function of measurement error and $\gamma,$ the photon loss rate. Measurement error is related to mean photon number as per Figs. \ref{fig:hetmsmterror},\ref{fig:ahdmsmt}, and is determined by the code parameters $N,K$ through $\bar{n}=(1/2)KN.$ The thresholds in the plots are determined for each code (fixed by $K,N$) by sweeping $\gamma$ for different code distances. 

Fig. \ref{fig:basicthresh} shows the threshold for a range of binomial codes against photon loss $\gamma=\kappa t_{\rm gate}$.  Binomial codes with varying values of $N$ are compared to the trivial encoding. The data point for a given value of $N$ and $K$ is plotted at the effective measurement error that is determine for this code in Sec.~\ref{1danalysis}. This allows a comparison of the threshold for the trivial encoding with a given qubit measurement error. In this initial example, Method 1 codes are used. The most striking feature of this plot is that the trivial encoding significantly outperforms the binomial codes, for all values of $N.$ Furthermore, increasing $N$ does not have an advantageous effect on the threshold, as might be expected. Finally, varying the binomial code parameter $K,$ and therefore varying the effective measurement error (for a code of fixed $N$) also does not appear to significantly change the threshold. 

These results can be understood as follows. Firstly examining Fig.~\ref{fig:hetmsmterror} we see that for heterodyne measurement of binomial codes, increasing $N$ causes a significant jump in the measurement error, for a given $K$. This is likely because each higher value of $N$ requires increased phase resolution for the phase measurement, going as $\pi/N$. Moreover the total photon loss probability is proportional to mean photon number $\bar{n}$, which in turn is proportional to $N$ for binomial codes. Finally the phase uncertainty of the heterodyne phase measurement is proportional to $1/\bar{n}$. Therefore, the improved tolerance to loss we expect to see when increasing $N$ for a binomial code is counteracted by the increase in measurement error and the increase in the total photon loss probability, meaning that we do not see an improvement in threshold. As $K$ decreases, the measurement error increases, meaning that the decreased mean photon number and therefore reduced probability of photon loss is counteracted by the increase in measurement error. This explains why the threshold values for the binomial codes of varying $K$ roughly lie on a straight line in Fig. \ref{fig:basicthresh}. 

These effects are strongly suggested by Fig. \ref{fig:normalisedthresh}, which plots the same data as Fig. \ref{fig:basicthresh}, except against $\gamma \bar{n}$ rather than $\gamma$. This effectively normalises the data against mean photon number.  We see that binomial codes of higher $N$ do better by this measure, and furthermore that binomial codes outperform the trivial encoding. Thus if we look at the thresholds against the probability of a single photon emission occuring $\gamma \bar{n}$ rather than against the probability for each photon to be lost $\gamma$ then indeed there is increased tolerance to photon loss errors for the higher $N$ codes, although it is the original thresholds measured against $\gamma$ that matter in practice.

We can improve upon the threshold results from Fig. \ref{fig:basicthresh} by using the Local Maximum Likelihood QSI rather than the Binning Algorithm QSI for single qubit measurements. Fig. \ref{fig:hetlocalml} shows that using the Local ML QSI, whilst still using heterodyne measurement, results in improved threshold values for codes of lower $N$. Local ML QSI takes into account an approximation of the local noise experienced by a qubit; the loss to which the qubit itself is subject, but also the dephasing errors that are propagated from loss errors on neighbouring qubits. Crucially, the dephasing errors are proportional to $\pi/N,$ which is just enough to cause a measurement error on neighbouring qubits. Therefore, the Local ML QSI can be expected to have a significant effect on threshold compared to the Binning Algorithm QSI.

We can further improve threshold results by using AHD measurement rather than heterodyne measurement. As shown in Fig. \ref{fig:ahdmsmt}, for AHD measurement, not only is the overall measurement error much lower than for heterodyne measurement, but also increasing $N$ increases the measurement error by a much smaller margin. The effects of this on threshold values are shown in Fig. \ref{fig:ahdlocalml}. We have significant overall improvements in threshold using AHD measurement, and in particular increasing $N$ corresponds to an increase in threshold, for a given value of measurement error. As increasing $N$ does not cause a large increase in measurement error for AHD measurement, the increase in $N$ reflects the increased capacity of a code to tolerate loss. 

The final improvement to threshold results we can make is by using modified MWPM decoding to improve MWPM. Modified MWPM decoding utilises the continuous variable nature of bosonic codes to weight error paths and improve the accuracy of the matching step of the MWPM algorithm. Fig. \ref{fig:dijkstraphase} shows that using modified MWPM decoding significantly increases threshold values relative to MWPM decoding. Interestingly, we see $N = 3$ and $N = 2$ outperforming the $N = 4$ code. For codes of higher $N$, it is more difficult to distinguish between $|\pm\rangle_{L}$ codewords as their angular separation in phase space is less and in addition they have higher photon numbers $\bar{n}$ leading to more photon loss events. Despite the increased loss tolerance of the high $N$ codes it appears that $N = 3$ is the optimal $N$ value when subject to these competing effects.

As we noted above the measurement errors resulting from imperfect phase measurement depend on the parameters of the binomial code. Higher values of $K$ offer enhanced phase resolution at the price of increased $\bar{n}$ as shown in Figs. \ref{fig:hetmsmterror},\ref{fig:ahdmsmt}. For this reason the best choice of code balances better phase resolution  with increased loss due to higher mean photon number. For both regular and modified MWPM decoding, this optimum  appears to be at a value of measurement error $~1\%$. This corresponds to an optimal choice of $K$ for each value of $N$.   This behaviour is quite distinct from what we saw in Fig. ~\ref{fig:basicthresh} where there was only a very weak dependence of the threshold on $K$. This is likely because the phase resolution in Method I is dominated by poor performance of heterodyne phase measurement which has phase uncertainty scaling like $1/\bar{n}$ relative to the AHD phase measurement which has uncertainty scaling like $1/\bar{n}^{3/2}$. This opens a window over which it is possible to improve performance by increasing phase resolution at the price of increased $\bar{n}$.

\subsubsection{Sub-threshold Behaviour}

While we have found that the thresholds against photon loss of the binomial codes are at best only comparable to the trivial encoding, this reflects the behaviour of these codes at high levels of photon loss. Far below the threshold the increased loss tolerance of the binomial codes could result in reduced overhead using binomial codes. This motivates us to look at the sub-threshold performance of these codes by investigate the scaling of logical errors with the distance of the planar code. 

Below the code threshold, it is well known \cite{rareeventssim} that the logical error rate of the code will scale as
\begin{align}
    p_{L}\propto e^{-\alpha d}
\end{align}
where $d$ is the code distance. Clearly, we may obtain $\alpha$ as the slope of a plot of $\mathrm{log}(p_{L})$ against $d.$
We quantify the performance of the codes in the subthreshold regime by this parameter  $\alpha.$ A steeper slope corresponds to a faster decay of the logical failure rate with increased code distance. This is desirable as it means that for a fixed target logical failure rate, a code with a larger $\alpha$ would require a smaller code size.

\begin{figure}[h]
\includegraphics[scale=0.4]{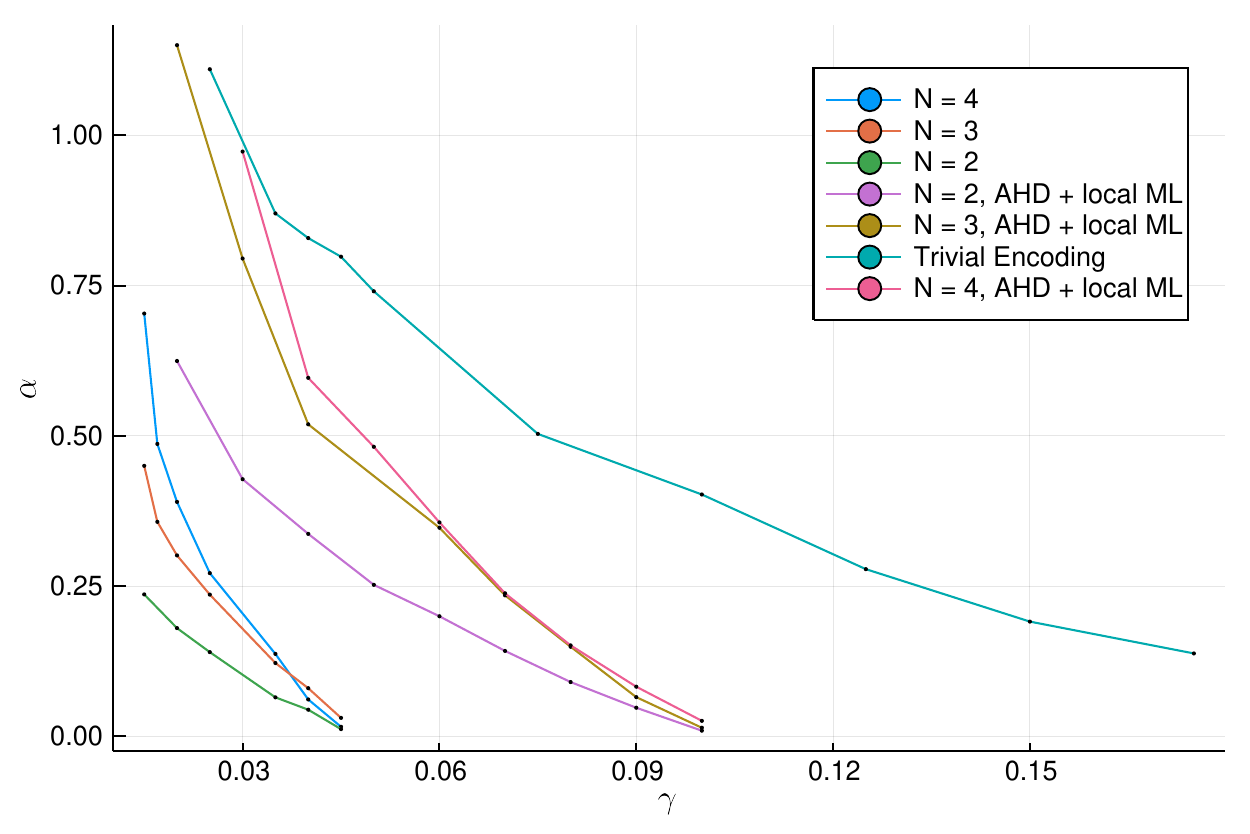}\\
\caption{Sub-threshold scaling parameter $\alpha$  against $\gamma$ for both Method 1 and Method 2 binomial codes and the trivial encoding. Higher $\alpha$ indicates more rapid reduction of logical errors with planar code distance. We see Method 2 codes significantly outperform compared to Method 1 codes in this regime. Furthermore, codes of higher $N$ perform better than their lower $N$ counterparts.}
\label{subthresh1}
\end{figure}

Fig. \ref{subthresh1} shows $\alpha$ vs $\gamma$ for binomial codes of $N = 2,3,4$ for both Method 1 and Method 2 Codes.
For Method 1 Codes, we see $\alpha$ increase as $\gamma$ decreases, as expected. We also see that higher $N$ corresponds to higher $\alpha,$ an effect which grows as $\gamma$ decreases. This is expected as codes with higher $N$ are better able to correct loss errors. In the sub-threshold regime where $\gamma$ is low, this property of higher $N$ codes is is evident as it is not outweighed by the detrimental effect of higher mean photon number. 
For Method 2 Codes, we see a similar pattern. In both cases, having higher $N$ appears to be advantageous. For a given value of $\gamma$, Method 2 codes have higher $\alpha$ values than Method 1 codes, leading us to the conclusion that Method 2 codes with a higher discrete rotational symmetry number are the best performing codes to use in the sub-threshold regime. Note that when Method 2 is used, the codes have a higher threshold compared to Method 1. Therefore, for a given $\gamma,$ Method 1 Codes will be at a lower fraction of the threshold value than Method 2 Codes.
The trivial encoding outperforms both Method 1 and Method 2 codes in the subthreshold regime. 

However again we note that the threshold of the trivial encoding is significantly higher than the binomial codes; therefore, for a given $\gamma$ the trivial encoding is at a lower proportion of threshold, so better performance is unsurprising. 

The comparison to the trivial encoding in the sub-threshold regime in particular should be interpreted as a helpful benchmark rather than a rigorous prediction of how the trivial encoding would actually perform in this regime. We have not attempted to use a realistic measurement model for the trivial encoding but rather ideal projective measurements with no measurement error. In contrast, we have subjected the binomial codes to realistic models of measurement. This effect is particularly important to note in the sub-threshold regime, where the strength of noise due to photon loss is low and therefore the impact of measurement error is more significant.

\begin{figure}[h]
\includegraphics[scale=0.4]{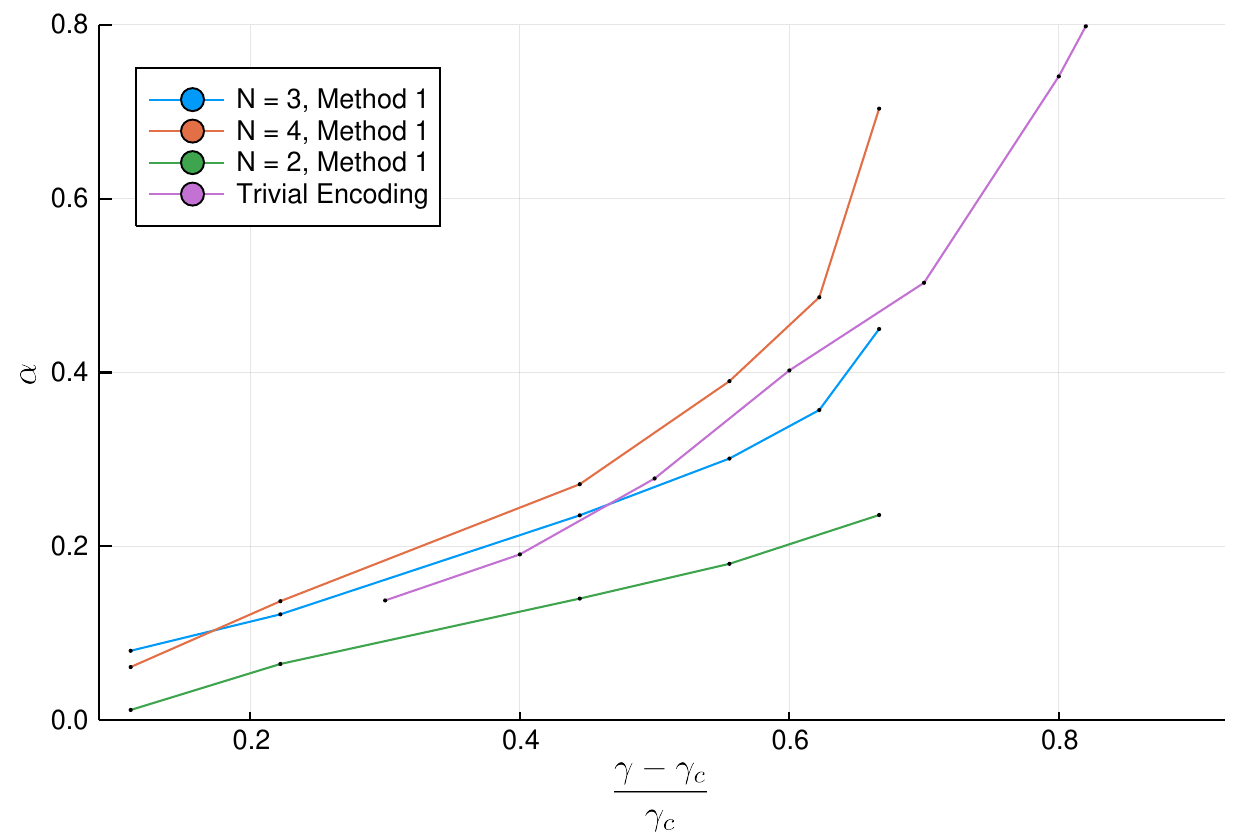}\\
\caption{Subthreshold scaling parameter $\alpha$ against $(\gamma_{c}-\gamma)/\gamma_{c}$ for Method 1 Codes. In this plot the threshold of all codes is at the left and the far-below threshold regime is to the right. The logical errors of higher $N$ codes are seen to drop more rapidly as $\gamma$ falls below threshold. }
\label{subthreshmethod1}
\end{figure}

\begin{figure}[h]
\includegraphics[scale=0.4]{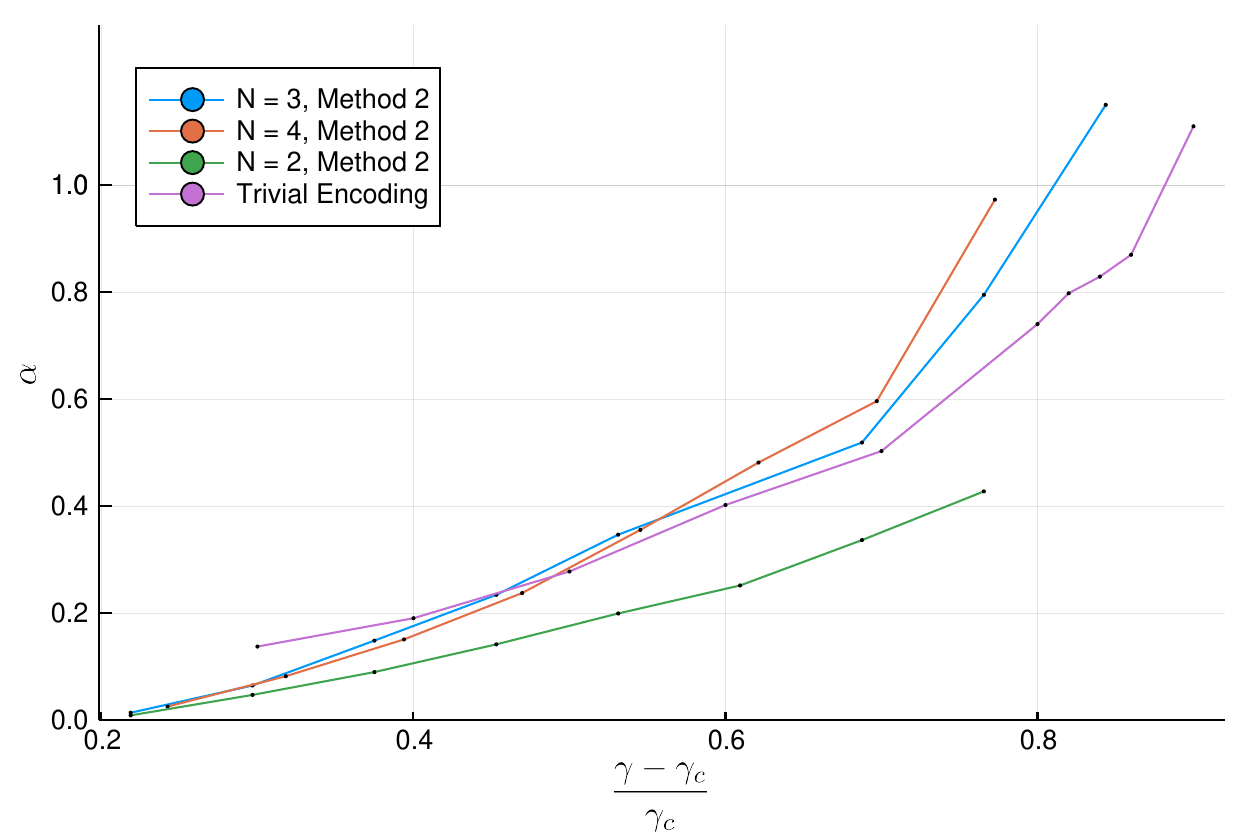}\\
\caption{Subthreshold scaling parameter $\alpha$ against $(\gamma_{c}-\gamma)/\gamma_{c}$ for Method 2 Codes. In this plot the threshold of all codes is at the left and the far-below threshold regime is to the right. The logical errors of higher $N$ codes are seen to drop more rapidly as $\gamma$ falls below threshold. The method 2 codes outperform method 1 codes.}
\label{subthreshmethod2}
\end{figure}

Figs. \ref{subthreshmethod1}\ref{subthreshmethod2} corrects for this by plotting $\alpha$ against $\gamma$ normalised by the threshold value, $\gamma_{c}.$
We are most interested in the right hand side of these plots, which is the region in which we are further away from $\gamma_{c},$ ie further below threshold. Sufficiently below threshold, Method 2 codes outperform their Method 1 counterparts with the same $N$ values. For $N=2,$ Method 2 codes outperform Method 1 for all fractions of threshold. For $N=3,4$ Method 2 codes do better lower than $30\%$ below threshold. Realistic quantum computers are expected to operate far below threshold and Method 2 codes have superior subthreshold scaling in this regime.

Furthermore, codes of higher $N$ outperform their lower $N$ counterparts sufficiently far below threshold. For Method 2 codes, below $50\%$ of threshold it is advantageous to have higher $N$. For Method 1 codes this is true below $80\%$ of threshold. For instance, for the $N = 4$ Method 1 code at  $\gamma = 0.035,$ we have $\alpha = 0.8,$ whilst for the $N = 4$ Method 2 code, $\alpha = 0.2$. This difference would scale up to a significant reduction in overhead in a realistic setting. Consider, for instance, using these codes to achieve a target logical error rate of $p_{L} = 10^{-10},$ which is standard for quantum chemistry algorithms \cite{quantchem}. For the Method 1 $N = 4$ code, this would require a code distance of 115, whilst for the Method 2 code it would require a code distance of 28. Given that the codes are 2D, this amounts to saving over 12000 qubits. 

Compared to the trivial encoding we see $N=4$ codes do better for Method 1 and $N=3,4$ codes do better for Method 2 when the location of threshold is taken into account. This reinforces the advantage obtained by using higher $N$ codes in the sub-threshold regime, especially for Method 2 codes. 

In this comparison where we have attempted to factor out the location of the threshold in order to study sub-threshold performance, note that that the location of the threshold will be a crucial consideration in practice. Namely, for codes with a higher threshold, it will be easier to access the sub-threshold regime. Overall the main point here is that codes with lower mean photon number can be highly advantageous, and that the choice of RSB code in a large-scale computation needs careful consideration. The best choice likely depends on details that we have not attempted to capture in this preliminary investigation.


\section{Discussion and Conclusion}\label{end}
In this paper, we have investigated a scheme for quantum computation in which the binomial code was concatenated with the surface code and realised by MBQC. We subjected this scheme to a realistic model of photon loss and compared several methods of phase measurement and qubit state inference. We also incorporated the soft information available from phase measurements into the decoding process by using a hybrid modified MWPM algorithm decoder. Using these techniques, we found phase diagrams of code thresholds as a function of loss, $\gamma$ and measurement error, $q$, for codes of varying orders of discrete rotational symmetry, $N$. Finally, we investigated the performance of these codes in the subthreshold regime. 

A key conclusion of our analysis is the impact of the quality of phase measurement on the performance of binomial codes. As shown in Fig. \ref{fig:hetmsmterror}, increasing the order of discrete rotational symmetry $N$, and therefore increasing the mean photon number, significantly increases the effective measurement error for heterodyne measurement. This counteracts the advantage of increased tolerance to photon loss possessed by codes of higher $N$. This effect can be clearly seen in Figs. \ref{fig:hetlocalml}, \ref{fig:basicthresh}, in which for a fixed $N,$ varying the measurement error by changing the binomial code parameter $K$ gives a roughly flat line in $\gamma.$ This affect arises because the probability of a photon loss event is roughly linear in $N$ while the phase resolution of the heterodyne measurement is roughly inversely proportional to $N$. The phase resolution of heterodyne measurement does not grow sufficiently with increased mean photon number to counteract the increased rate of photon loss events in that limit.

It is therefore necessary to use an adaptive homodyne phase measurement to realise the full power of the binomial codes. The improved phase resolution of adaptive homotdyne schemes enables the increased loss tolerance of higher $N$ codes to `win out,' resulting in higher thresholds against loss. 

The other main performance improvement we saw came from using the modified algorithm decoding rather than unweighted MWPM decoding. By using the soft information of the continuous variable measurement outcomes, we were able to achieve a threshold against loss of $~20\%$ , compared to $~14\%$ using MWPM. This is not only a significant improvement, but is objectively a high threshold against loss.

An important consideration is our comparison of the performance of binomial codes to the trivial encoding. We find that even our best scheme for phase measurement and quantum state inference results in thresholds that are at best comparable to those of the trivial encoding with some measurement error probability. The trivial encoding only has a maximum of a single photon per qubit, and therefore performs relatively well against loss because it has a very low mean photon number. We believe that the conclusion that should be drawn from this is that the which error correcting coding is best in a given situation will depend on details of the noise model in a given physical system that are beyond the scope of this work. When modelling the trivial encoding we assumed ideal measurements and added the measurement error as classical noise on the measurement outcome. A more accurate comparison would simulate realistic measurement for the trivial encoding. Whether or not the trivial encoding does actually outperform the binomial code would depend on the noise of such a detailed implementation. One way of making a comparison between the trivial and binomial codes in these simulations is to note that the threshold against loss that we found corresponds to a trivial encoding with a measurement error rate of around $1\%$.

Finally we studied the supression of logical errors below threshold. In this case we do see that the binomial codes are able to suppress logical errors more rapidly with increasing $N$ when the photon loss rate is low enough. 

Our analysis has shown that for the binomial code concatenated with the surface code, the performance is highly dependent upon accurate phase measurement, qubit state inference and decoding. All of these features need to be well chosen for the apparent advantage of binomial codes in tolerating photon loss errors to be realised in practice. 

\section*{Acknowledgements}

We thank Ben Brown for his contributions to the initial stages of this project. We acknowledge support from Australian Research Council via the Centre of Excellence in Engineered Quantum Systems (CE170100009).
We acknowledge the traditional owners of the land on which this work was undertaken at the University of Sydney, the Gadigal people of the Eora Nation.

\onecolumngrid

\appendix

\section{Quantum Trajectory Simulation of Loss}
\label{quanttraj}
\subsection{Justification of independent sampling of cluster qubits}
We aim to produce the state corresponding to the graph graph $G = (V,E)$ which has $M$ vertices total. This can be produced by applying a $\textsc{CROT}$ gate to all pairs of qubits that share an edge of $G$: 
\begin{align}
    U =\prod_{e\in E} \textsc{CROT}_{e_1,e_2} = e^{-iH_{\textsc{CROT}}t_{\rm gate}}
\end{align}
where 
\begin{align}
    H_{\textsc{CROT}} &= -\Omega\sum_{e\in E}\hat{n}_{e_1}\otimes\hat{n}_{e_2}
\end{align}
and $t_{\rm gate} = \pi/N^2\Omega.$ 

We aim to simulate photon losses occurring at random times during the implementation of these gates using a quantum trajectories approach, as per \cite{howard}. Having found an analytic expression for these quantum trajectories we show by analysing  the norm of the full evolution of the cluster that loss times can be sampled independently for each qubit of the cluster. 

According to quantum trajectory theory the evolution of the quantum state between photon emissions is given by the non-Hermitian Hamiltonian
\begin{align}
    \frac{d}{dt}|\tilde{\psi}(t)\rangle &=-iH_{\rm eff}|\tilde{\psi}(t)\rangle\nonumber \\
    &=\left(-i\Omega\sum_{e\in G}\hat{n}_{e_1}\otimes\hat{n}_{e_2} -\frac{\kappa}{2}\sum_{a=1}^{M}\hat{n}_{a}\right)|\tilde{\psi}(t)\rangle
\end{align}
where the tilde indicates that the state is unnormalised.\\

Suppose that $k$ photons in total have been emitted during the gate time $t_{\rm gate}$ for the gate $\textsc{CROT}$ gate. The $j$th photon emission removes a photon from the mode $\alpha_j$ with $j=1,\ldots,k$. The time between the $j$th and $j+1$th photon emission is $\Delta_j$. ($\Delta_k=t_{\rm gate}-t_{k}$ where $t_k$ is the final photon emission time).  

As in the main text the vector $\mathbf{t}_a$ is a temporally ordered list of photon emission times for mode $a$. The number of emissions for mode $a$ is $j_a=|\mathbf{t}_a|$. The full set of emission times is recorded in the array $\mathbf{t}= \{\mathbf{t}_1,\ldots, \mathbf{t}_M\}$. We have $k=\sum_{a}j_a$. 
%Let $\sum_{\beta_{i}} = \sum_{\beta_{i}\in\mathcal{N}(\alpha_{i})}$ denote the sum over modes index by $\beta_{i}$ which are entangled to the mode $\alpha_{i}$ by the $\textsc{CROT}$. 

Trajectory theory states that the non-Hermitian time evolution operator corresponding to the noisy gate with the specified photon emission times is \begin{equation}
    \tilde{U}_{\mathbf{t}}=\sqrt{\kappa}^k e^{-iH_{\rm eff}\Delta_{k}}a_{\alpha_{k}}...e^{-iH_{\rm eff}\Delta_{1}}a_{\alpha_{1}}e^{-iH_{\rm eff}t_{1}}.
\end{equation}
We will make use of the following identities \cite{arne} to simplify this expression,
\begin{align}
    e^{c\hat{n}}\hat{a}^{k} &= e^{-ck}\hat{a}^{k}e^{c\hat{n}}\\
    e^{ic\hat{n}_{1}\otimes\hat{n}_{2}}\hat{a}_{1}^{k}&=e^{-ick\hat{n}_{2}}\hat{a}_{1}^{k}e^{ic\hat{n}_{1}\otimes\hat{n}_{2}}.
\end{align}

One can show using induction that the time evolution operator corresponding to this sequence of photon loss events is
\begin{align}\label{identityequation}
\tilde{U}_{\mathbf{t}}
    &=\sqrt{\kappa}^k e^{\kappa\sum_{m=1}^{k}\Delta_{m}m/2}e^{i\Omega\sum_{m=1}^{k}(\sum_{l=m}^k\Delta_{l})\sum_{b\in\mathcal{N}(\alpha_m)}\hat{n}_{b}}\left(\prod_{m=1}^{k}\hat{a}_{\alpha_{m}}\right)e^{-iH_{\rm eff}t_{\rm gate}}\nonumber \\
    &=\sqrt{\kappa}^k \prod_{a=1}^{M}\left[e^{\kappa \tau(\mathbf{t}_a)/2}\left(\prod_{b\in\mathcal{N}(a)}e^{i\Omega\sum_{t_{m}\in\mathbf{t}_{b}}(t_{m+1}-t_{m})m\hat{n}_{a}}\right)\hat{a}_{a}^{j_a}\right]e^{-iH_{\rm eff}t_{\rm gate}}\nonumber \\
    %&=\sqrt{\kappa}^k e^{\frac{\kappa}{2}\sum_{m=1}^{k}\Delta_{m}m}\left(\prod_{a=1}^M e^{i\Omega\sum_{m:t_m\in\mathbf{t}_j}(t_{m+1}-t_{m})\epsilon_{j_{m}}\hat{\mathbf{n}}_{a}}\right)\left(\prod_{a=1}^{M}\hat{a}_{a}^{j_a}\right)e^{-iH_{\rm eff}t_{gate}}\nonumber \\
    &= 
    %\left(\prod_{a=1}^M e^{i\Omega\sum_{t_m\in\mathbf{t}_j}\Delta_{m}\epsilon_{j_{m}}j_a}\right)
    \left(\prod_{a=1}^M \hat{E}_{a,\mathbf{t}}\right)U.
\end{align}
The operators $\hat{E}_{a,\mathbf{t}}$ are as defined in the main text. We have used the fact that $\sum_{t_m\in\mathbf{t}_a}(t_{m+1}-t_{m})m=j_at_{\rm gate}-\sum_{t_m\in \mathbf{t}_a}t_m=\tau(\mathbf{t}_a)$.

This expression gives the error process as the product of ideal $\textsc{CROT}$ gates followed by photon losses on each qubit. We used this representation of the errors in the main text and also in our simulations. However for the simulations it is necessary to be able to sample accurately from the distribution over photon loss events $\mathbf{t}$. To compute the probability density for this pattern of emissions $\mathbf{t}$ it is helpful to also be able to write the noise operator as photon emissions followed by the ideal gate. We can rearrange the expression above as follows \begin{align}\tilde{U}_{\mathbf{t}} =\sqrt{\kappa}^k e^{\kappa \sum_{a=1}^M \tau(\mathbf{t}_a)/2}U \left[\prod_{a=1}^M\left(\prod_{b\in\mathcal{N}(a)} e^{i\Omega\left[\tau(\mathbf{t}_b)-j_bt_{\rm gate}\right]\hat{{n}}_{a}}\right)\right]\left(\prod_{a=1}^{M}\hat{a}_{a}^{j_a}\right) e^{-\kappa t_{\rm gate}\sum_a^M\hat{n}_a/2}\end{align}

We now consider the inner product
\begin{align}\label{eq19}
\langle +|^{\otimes M}\tilde{U}_{\mathbf{t}}^\dagger\tilde{U}_{\mathbf{t}}|+\rangle ^{\otimes M}&= \kappa^k e^{\kappa\sum_{a=1}^M \tau(\mathbf{t}_a)}\langle +|^{\otimes M}e^{-\kappa t_{\rm gate}\sum_a^M\hat{n}_a/2}\left(\prod_{a=1}^{M}\hat{a}^{\dagger j_a}_{a}\right) \left(\prod_{a=1}^{M}\hat{a}_{a}^{j_a}\right) e^{-\kappa t_{\rm gate}\sum_a^M\hat{n}_a/2}|+\rangle ^{\otimes M} \nonumber \\ &= \kappa^k e^{\kappa\sum_{a=1}^M \tau(\mathbf{t}_a)}\langle +|^{\otimes M}e^{-\kappa t_{\rm gate}\sum_a^M\hat{n}_a/2}\left[\prod_{a=1}^{M}:\left(\hat{a}_a^\dagger \hat{a}_{a}\right)^{j_a}:\right]  e^{-\kappa t_{\rm gate}\sum_a^M\hat{n}_a/2}|+\rangle ^{\otimes M}\nonumber \\
&= \prod_{a=1}^M\kappa^{j_a} e^{\kappa\tau(\mathbf{t}_a)}\langle +|e^{-\kappa t_{\rm gate}\hat{n}_a/2}:(\hat{a}^\dagger_{a}\hat{a}_{a})^{j_a}:  e^{-\kappa t_{\rm gate}\hat{n}_a/2}|+\rangle 
\end{align}

Where $::$ indicates the normally ordered product of operators and so
\begin{align}
    :(\hat{a}^{\dagger}\hat{a})^{k}:&=(\hat{a}^{\dagger})^{k}\hat{a}^{k}
\end{align}
Eq. \ref{eq19} shows that the probability for obtaining a given pattern of photon emissions is a simple product distribution over the modes. This greatly simplifies the task of drawing random samples $\mathbf{t}$  of photon emission events during the $\textsc{CROT}$ gate. Moreover it is clear that the probability distribution has no dependence on $\Omega$. A simple calculation shows that in fact the photon emission probabilities are the same as for uncoupled modes experiencing photon loss at rate $\kappa$ for a time $t_{\rm gate}$ and no other dynamics.
This justifies the approach of independently sampling loss times for the cluster qubits that is described in detail in the following subsection. 

\subsection{Sampling Algorithm}\label{dist}
Given the result on the statistic of the photon emissions, we choose to simplify the numerical recipe by iterating across each qubit of the cluster, and sampling from each qubit independently as follows:\\
1. Generate a uniform random number $R$.\\
2. Iteratively solve the Schrodinger equation
\begin{align}
    \frac{d}{dt}|\tilde{\psi}(t)\rangle &= -\frac{\kappa}{2}\hat{n}|\tilde{\psi}(t)\rangle.
\end{align}
3. Do step 2 up until time $T$ at which
\begin{align}
    \langle\tilde{\psi}(T)|\tilde{\psi}(T)\rangle &= \langle+|e^{-\hat{n}_{\mu}\kappa T}|+\rangle< R.\nonumber
\end{align}
Apply the jump operator $\hat{c}_{\mu} = \sqrt{\kappa} \hat{a}_{\mu}$, renormalise the state.\\
4. Repeat steps 1-3 until $t = t_{\rm gate}.$\\

The output of this sampling is a particular set of photon emissions $\mathbf{t}$. The noisy state is then obtained by applying the noise operator $\hat{E}_{a,\mathbf{t}}$ to each qubit $a$.

\section{POVMs for AHD Measurement} 
\label{ahdappendix}
We follow the presentation of \cite{arnerecent} in giving the theoretical details of AHD POVMs as follows
\begin{align}
    H_{mn}^{AHD} &= \sum_{p=0}^{\left\lfloor{\frac{m}{2}}\right\rfloor}\sum_{q=0}^{\left\lfloor{\frac{n}{2}}\right\rfloor}\gamma_{m,p}\gamma_{n,q}C^{(m,n)}_{p,q}
    \end{align}
    where
    \begin{align}
    \gamma_{m,p} &= \frac{\sqrt{m!}}{2^{p}(m-2p)!p!}\\
    C^{(n,m)}_{p,q}&= \sum_{l=0}^{\infty}\sum_{l'=0}^{\infty}\binom{\frac{n-m}{2}}{l}\binom{\frac{m-n}{2}}{l'}M_{p+l,q+l'}
    \end{align}
    and 
    \begin{align}
    \binom{\alpha}{n}&= \prod_{k=1}^{n}\frac{\alpha-k+1}{k}
\end{align}
are generalised binomial coefficients. The $M_{m,n}$ are recursively defined
\begin{align}
    M_{m,n} &=\frac{nM_{n-1,m}+mM_{n,m-1}}{2(n-m)^{2}+n+m}\\
    M_{n,0} &= M_{0,n} = \frac{1}{(2n+1)!!}.
\end{align}

\section{X Basis Pauli Twirl}
\label{Xbasispaulitwirl}

In the following we will show the invariance of the probability distribution of qubit measurement outcomes under an X-basis Pauli twirl in the case $\gamma=0$.

We start by defining the measurement operators $M_\pm$ as described in the main text in the limit of no photon loss. Given a phase measurement $\hat{F}(\phi)$ the POVM elements for qubit measurement are as follows
\begin{align}\label{phasepovmeq}
    M_{\pm,a,\mathbf{t}} = \int_{\phi\rightarrow\pm}d\phi\hat{F}(\phi)
\end{align}
The integral over $\phi$ is determined by the QSI technique resulting in the measurement outcome either being binned as $+1$ or $-1$. We will restrict our attention here only to a pure binning QSI.

We will show that \begin{align}\label{initialstatement}
\langle+|M_\pm|-\rangle=0=\langle-|M_\pm|+\rangle
\end{align}
where $|\pm\rangle$ are the $X$-eigenstates of our RSB code.

If this we perform this measurement on a qubit $a$ that is part of a RSB code cluster state $\rho$ then this property is sufficient to ensure that the conditioned state satisfies \begin{equation}\rho_{a,\pm}=\mathrm{Tr}_a[M_{a,\pm} \rho]=\mathrm{Tr}_a[M_{a,\pm}\mathcal{P}_{X_a}(\rho) ].\end{equation} This identity means performing an $X$-basis twirl on qubit $a$ does not affect either the probability of the measurement outcome or the post-measurement state.

Before making this calculation we recall the following definitions. 
An arbitrary phase measurement can be represented by the general POVM \cite{ahdmain}
\begin{align}
    \hat{F}(\phi) &= \frac{1}{2\pi}\sum_{n,m=0}^{\infty}e^{i(m-n)\phi}H_{mn}|m\rangle\langle n|
\end{align}
where $H_{mn}$ is a Hermitian matrix with real positive entries, $H_{mm}=1$ for all $m$ and $\phi\in[0,2\pi].$ 
And for a rotational symmetric bosonic code \cite{arne}
\begin{align}
    |+\rangle &=\frac{1}{\sqrt{2}}\sum_{k=0}f_{k}|kN\rangle\nonumber\\
    |-\rangle &=\frac{1}{\sqrt{2}}\sum_{k=0}f_{k}(-1)^{k}|kN\rangle
\end{align} where the $f_k$ are real coefficients. The $f_k$ satisfy the following normalisation conditions \begin{equation*}
    \sum_{k\ \mathrm{even}}f_k^2=1=\sum_{k\ \mathrm{odd}}f_k^2.
\end{equation*}

Suppose that we are using the binning algorithm QSI. There are $2N$ binning regions centered at $\pi\cdot l/N$ for $l\in\{0,1,...,2N-1\}.$ Each binning region spans the angles $\pi(l-1/2)/N\rightarrow \pi(l+1/2)/N$ where even $l$ corresponds to $+1$ bins and odd $l$ corresponds to $-1$ bins. The POVM for `measuring a $\pm1$ X basis outcome' is therefore given by the sum over binning regions for even (odd) $l$, respectively. We will now integrate the term with $\phi$ dependence from Eq. \ref{phasepovmeq}, $e^{i(k-k')N\phi},$ over the binning regions
\begin{align}
    \int_{\phi\rightarrow +1}d\phi e^{i[(k-k')N]\phi}&=\sum_{l=0\textrm{, $l$ even}}^{2N-1}\int_{\pi(l-1/2)/N}^{\pi(l+1/2)/N}e^{i(k-k')N\phi}d\phi\\
    \int_{\phi\rightarrow -1}d\phi e^{i[(k-k')N]\phi}&=\sum_{l=0\textrm{, $l$ odd}}^{2N-1}\int_{\pi(l-1/2)/N}^{\pi(l+1/2)/N}e^{i(k-k')N\phi}d\phi.
\end{align}

Consider the case $k\neq k'$
\begin{align}\label{intbins}
    c_{k,k',l}=&\int_{\pi(l-1/2)/N}^{\pi(l+1/2)/N}e^{i(k-k')N\phi}d\phi \nonumber \\  =&
    \frac{2}{(k-k')N}\cos[(k-k')\pi l]\sin[(k-k')\pi/2]
\end{align}
This is zero whenever $k-k'$ is even, and is an even function of $k-k'$ when that number is odd. 

In the case $k=k'$ we have 
\begin{align}\label{intbins}
    c_{k,k,l}=\int_{\pi(l-1/2)/N}^{\pi(l+1/2)/N}e^{i(k-k)N\phi}d\phi &=\frac{\pi}{N}
\end{align}

Now we can find \begin{align}
   \langle +| M_+ |-\rangle = &\frac{1}{4\pi}\sum_{l=0,\mathrm{even}}^{2N-1}\sum_{k,k'}\left(\int_{\pi(l-1/2)/N}^{\pi(l+1/2)/N}e^{i(k-k')N\phi}d\phi \right) f_kf_{k'}(-1)^{k'}H_{kN,k'N} \nonumber \\ 
    = &\frac{1}{4\pi}\sum_{l=0,\mathrm{even}}^{2N-1}\sum_{k,k'}c_{k,k',l} f_kf_{k'}(-1)^{k'}H_{kN,k'N} \nonumber \\ 
    = &\frac{1}{4}\sum_k f_k^2(-1)^kH_{kN,kN}+\frac{1}{4\pi}\sum_{l=0,\mathrm{even}}^{2N-1}\sum_{k\neq k'}c_{k,k',l} f_kf_{k'}(-1)^{k'}H_{kN,k'N} 
\end{align}
The first term here is zero because $H_{mm}=1$ for all $m$ and due to the normalisation conditions on the $f_k$. The final term is also zero. When $k-k'$ is even the coefficients $c_{k,k',l}$ are zero. In the case that  $k-k'$ is odd the terms with $k,k'$ and $k',k$ cancel pairwise. The factor in the summand $c_{k,k',l}f_kf_{k'}H_{kN,k'N}$ does not change if we swap $k$ and $k'$ while the factor $(-1)^{k'}$ changes sign. This is because if $k-k'$ is odd then one and only one of $k,k'$ are odd. Thus we have $\langle +|M_+|-\rangle=0$ as required.

This gives the result for $M_+$, the argument for $M_-$ is exactly the same.

\twocolumngrid

\bibliography{paperdraftmain}

\end{document}